\documentclass[floats,floatfix,showpacs,amssymb,prd,twocolumn,superscriptaddress,nofootinbib]{revtex4-1}

\usepackage{graphicx, epsf, epsfig, amssymb}
\usepackage{bm}
\usepackage{longtable}
\usepackage[usenames,dvipsnames]{xcolor}
\usepackage{color}
\usepackage[breaklinks]{hyperref}
\usepackage{amsfonts,amsmath,amssymb,mathrsfs}

\newcommand{\tn}{\textnormal}

\def\be{\begin{equation}}
\def\ee{\end{equation}}
\def\beq{\begin{eqnarray}}
\def\eeq{\end{eqnarray}}

\begin{document}
 
\title{
Constraining the equation of state of nuclear matter with gravitational\\ 
wave observations: Tidal deformability and tidal disruption}

\author{Andrea Maselli}

\affiliation{Dipartimento di Fisica, Universit\`a di Roma ``La Sapienza'' \& Sezione, INFN Roma1, P.A. Moro 5, 00185, Roma, Italy.}

\author{Leonardo Gualtieri}
\affiliation{Dipartimento di Fisica, Universit\`a di Roma ``La Sapienza'' \& Sezione, INFN Roma1, P.A. Moro 5, 00185, Roma, Italy.}

\author{Valeria Ferrari}
\affiliation{Dipartimento di Fisica, Universit\`a di Roma ``La Sapienza'' \& Sezione, INFN Roma1, P.A. Moro 5, 00185, Roma, Italy.}


\date{\today}

\begin{abstract}
We study how to extract information on the neutron star equation of
state from the gravitational wave signal emitted during the coalescence
of a binary system composed by two neutron stars or a neutron star and a
black hole. We use Post-Newtonian templates which include the tidal
deformability parameter and, when tidal disruption
occurs before merger, a frequency cut-off.
Assuming that this signal is detected by Advanced LIGO/Virgo or ET,
we evaluate the uncertainties on these parameters using different data
analysis strategies based on the Fisher matrix approach, and on
recently obtained analytical fits of the relevant quantities.
We find that the tidal deformability is more
effective than the stellar compactness to discriminate among different
possible equations of state.
\end{abstract}

\maketitle

\section{Introduction}\label{sec:intro}
Black hole (BH)-neutron star (NS) mergers are promising sources of gravitational
waves (GWs) to be detected by interferometric detectors of second (AdvLIGO/Virgo
\cite{LIGOVirgo}) and third (ET, \cite{ET}) generation. The detection of GW
signals from these processes will provide valuable information on the NS
internal structure, which would be impossible to obtain otherwise.  For this
reason, in recent years much effort has been devoted to model the signal emitted
during the latest phases of the inspiralling of NS-NS and NS-BH binaries, when
the imprint of the equation of state (EoS) on the signal is more pronounced (see
e.g.
\cite{FH08,H08,Val200,FGP10,HLLR10,KST10,DFKOT10,PTR11,LKSBF11,DNV12,Ral13}).
However, the best strategy to extract as much of information as possible on the
NS EoS from a detected signal is still under debate.  Addressing this question
is the main scope of this work.  We shall focus on the latest inspiralling, when
NS deformations can be large, and we shall use two quantities which encode
information on the NS EoS, to compare different strategies: the tidal
deformability $\lambda$ (which will be defined in Section~\ref{sec:model}) and
the stellar compactness ${\cal{C}}=M_{\tn{NS}}/R_{\tn{NS}}$, where $M_{\tn{NS}}$
is the NS gravitational mass and $R_{\tn{NS}}$ its radius.

If the companion is a black hole, under appropriate conditions a NS can be
disrupted by the tidal interaction before being swallowed.  In this case the
GW-signal exhibits a frequency cut-off $f_{\tn{cut}}$ which also depends on the
EoS.  Most of the articles cited above include either $\lambda$ (or ${\cal{C}}$)
or $f_{\tn{cut}}$ in the model; in this paper, we shall include both parameters
in the signal template and, using recently proposed analytic fits of
${\cal{C}}(\lambda)$ \cite{MCFGP13} and $f_{\tn{cut}}({\cal{C}})$ \cite{KST10},
and a Fisher matrix approach, we shall evaluate whether a joint analysis can
improve the possibility to gain information on the NS EoS with respect to an
approach which includes only one of these quantities.  It should be stressed
that tidal disruption in BH-NS or NS-NS binaries is of particular interest,
since they are have been invoked as possible engines of Short Gamma-Ray Bursts
\cite{NPP92} (for a review on the subject, see e.g. \cite{BBM13}).

In addition, we shall discuss which is the most useful quantity to be used in
order to constrain the EoS of matter in the NS interior.  In the literature, it
is often assumed that the most valuable information that GW physicists (or
astrophysicists) can provide to nuclear physicists is the value of the NS radius
$R_{\tn{NS}}$ or, equivalently, the star compactness ${\cal C}$ \cite{LP07}. A
possibility is to estimate $\lambda$ from a detected GW signal and then derive
$R_{\tn{NS}}$ from it \cite{YY13,MCFGP13}.  On the other hand, $\lambda$ itself
could be used to constrain the NS EoS \cite{HLLR10,DNV12}.  We shall compare
these two approaches, assessing their ability to discriminate among different
EoS.

In this paper we shall consider the advanced GW detectors AdvLIGO/Virgo, and
third generation detectors like ET. Since ET is expected to detect GW signals
from coalescing binaries up to 2 Gpc, the cosmological redshift will be
consistently included in the template waveforms.

The plan of the paper is the following.  In Section \ref{sec:model} we briefly
introduce the tidal deformability $\lambda$ and the cut-off frequency and in
Section \ref{sec:template} we show how these parameters can be included in a
waveform template.  In Section \ref{sec:strategies} 
we discuss how $\lambda$, $f_{\tn{cut}}$ and ${\cal{C}}$ can be used to extract
information on the NS EoS. In Section \ref{sec:lambdaC} we discuss whether the
tidal deformability has to be preferred to the stellar compactness as a
parameter to discriminate among different EoS.  In Section \ref{sec:concl} we
draw our conclusions.  Finally, in Appendix \ref{app:Fisher} we sketch the basic
elements of parameter estimation theory. We remark that Section
\ref{sec:strategies} refers to BH-NS binary systems, while in the rest of the
paper we consider both BH-NS and NS-NS binaries.

\section{Modeling tidal interactions in BH-NS and NS-NS
binaries}\label{sec:model}
In this Section we briefly introduce the parameters 
which depend on the equation of state of matter in the NS interior,
and which appear in the late inspiralling GW signal: 
the Love number, describing the star deformability, and the cut-off 
frequency which appears in the emitted signal when a NS is tidally
disrupted by a companion BH before merging.
\subsection{The Love number}
NS tidal deformations can be described in terms of a set of parameters,
the Love numbers \cite{FH08,H08}, which relate the mass multipole
moments of the deformed star with the external tidal multipole moments.
The quantity which encodes most of the information on stellar
deformation is  the quadrupole Love number $k_2$, given by
\begin{equation} 
Q_{ij}=-\frac{2}{3}k_2 R_{\tn{NS}}^5  C_{ij}\label{defk2} 
\end{equation} 
where $Q_{ij}$ is the star traceless quadrupole tensor, and
$C_{ij}=e^\alpha_{(0)}e^\beta_{(i)}e^\gamma_{(0)}e^\delta_{(j)}R_{\alpha\beta\gamma\delta}$
is the tidal tensor of the gravitational field ($e^\alpha_{(\mu)}$ is the
parallel transported tetrad attached to the deformed star, and
$R_{\alpha\beta\gamma\delta}$ is the Riemann tensor).  The ratio between the
quadrupole and the tidal tensors,
\begin{equation}
\lambda=\frac{2}{3}k_2 R_{\tn{NS}}^5\label{deflambda}\,, 
\end{equation}
is the NS {\it tidal deformability}.

$k_2$ and $\lambda$ can be computed by studying the quadrupolar, stationary
perturbations induced by a test tidal field acting on the NS
\cite{H08,DN09,BP09}. It has recently been shown \cite{YY13} that the Love
number - or, more precisely, the tidal deformability - is related to the
momentum of inertia of the star and to its rotation-induced quadrupole moment by
the so-called ``I-Love-Q relations'', which are almost independent of the EoS
and of the NS mass; a similar universal relation can be found between the tidal
deformability and the NS compactness ${\cal C}$ \cite{MCFGP13}.

Tidal interactions affect the gravitational signal emitted by coalescing compact
binaries, and these effects are currently included in the gravitational
waveforms within approximation schemes, as the Post-Newtonian expansion (PN), or
the Effective One Body (EOB) formalism. In these frameworks it has been shown
\cite{VFH11,BDF12} that the leading contribution is given by an extra term in
the phase of the signal, which is proportional to the tidal
deformability. Therefore, the detection of a signal emitted in a NS-BH or a
NS-NS coalescence can, in principle, allow to determine $\lambda$ or $k_2$. The
accuracy with which second and third generation detectors will be able to
measure $\lambda$ has recently been estimated through a data analysis based on
Fisher matrix or Markov Chain Monte Carlo techniques (see App.~\ref{app:Fisher}
and \cite{CF94} for a general discussion on the Fisher matrix formalism; see
\cite{BSD96} for the Markov Chain Monte Carlo approach). Using the EOB
formalism and a large class of EoS, it has been shown that $\lambda$ can be
measured by advanced detectors if the signal emitted by a coalescing NS-NS
binary is detected with signal-to-noise ratio (SNR) $\rho=16$ \cite{DNV12}.  In
the case of BH-NS binaries, a similar estimate has been carried out using
phenomenological waveforms obtained matching PN templates with the outcome of
fully relativistic simulations, finding that the tidal deformability can be
extracted for sources whose mass ratio $q=M_{\tn{BH}}/M_{\tn{NS}}$ (where
$M_{\tn{BH}}$ is the BH mass) is $q=2,3$, at $d=100$ Mpc with 10\%-40\% accuracy
by AdvLIGO/Virgo, and with an order of magnitude better accuracy by the Einstein
Telescope \cite{LKSBF11}. Markov Chain Monte Carlo approaches have been
employed in \cite{DABV13}, showing that, for NS-NS binaries, few tens of
detections by advanced interferometers will be required to strongly constrain
$\lambda$.

\subsection{The frequency cut-off}\label{sec:cutoff}

In the coalescence of a BH-NS binary system, the NS can be tidally disrupted
before plunging into the BH. This phenomenon can occur if the star is very
deformable, if the BH is rapidly spinning, or if the mass-ratio is very small
(see e.g. \cite{SKYT09,KOST11,KST10,PTR11}). Otherwise, the NS behaves nearly as
a point particle until the very last stages of the coalescence, and it is
swallowed by the BH without being disrupted.

When the NS is tidally disrupted, the gravitational signal exhibits a clear
signature: the waveform amplitude steeply drops down at a {\it cut-off}
frequency $f_{\textnormal{cut}}$ \cite{Val200} (see also \cite{BBM13} for a
general description of the process).  So far, studies of $f_{\textnormal{cut}}$
for BH-NS binaries have been carried out with either semi-analytical approaches
\cite{Val200,FGP10} or fully numerical simulations
\cite{SKYT09,KST10,KOST11}. The first proposal of using a measure of
$f_\tn{cut}$ to determine $R_{\tn{NS}}$ goes back to more than a decade ago
\cite{Val200}; in that article a relation, based on simplifying assumptions and
derived in a Newtonian framework, was proposed to connect $f_\tn{cut}$ with
$R_{\tn{NS}}$.  A much more accurate relation between $f_\tn{cut}$ and the
compactness ${\cal C}$ (and then $R_{\tn{NS}}$) has recently been proposed
\cite{KST10}, based on the results of fully relativistic simulations of BH-NS
coalescences:
\begin{equation}\label{fcutC}
\ln(f_\textnormal{cut}m)=(3.87\pm 0.12)\ln {\cal C}+(4.03\pm 0.22)
\end{equation}
where $m\equiv M_{\tn{NS}}+M_{\tn{BH}}$.  These simulations have been performed
assuming that the BH is non-rotating, and that the mass-ratio of the binary is
$q=2$, therefore the fit (\ref{fcutC}) can only be trusted under these
assumptions. In the next two Sections we shall exploit the fit (\ref{fcutC});
therefore, we shall focus on BH-NS binaries with zero BH spin and $q=2$.

Realistic BH-NS binary systems are thought to have larger values of the mass
ratio and a large BH rotation rate \cite{N05}. The analysis carried out in the
present paper can easily be extended to such configurations, once fully
relativistic simulations of these systems will be available.

\section{The gravitational wave templates}\label{sec:template}
Post-Newtonian GW templates which include finite size effects, accurately
describe the evolution and the gravitational emission of BH-NS and NS-NS
binaries up to the last orbits before the merger
\cite{HKS13,RLG13,BDGNR,BTB12,BNTB12}.  However, they do not encode information
about NS tidal disruption which may possible occur. This can be done, as a first
approximation, including in the gravitational template a cut-off at the
frequency $f_\tn{cut}$; this value will be treated as one of the parameters of
the template.  We shall model the inspiral waveform as follows:
\begin{align}\label{hfit}
\bar{h}_{\tn{PN}}(f)=\begin{cases}
h_{\textnormal{3PN}} & f<f_{\tn{cut}}\\
h_{\textnormal{3PN}}\times \Theta(f,f_{\tn{cut}}) & f_{\tn{cut}}\leq f\leq 2f_{\tn{cut}}\\
0 & f>2f_{\tn{cut}}\\
\end{cases}
\end{align}
where 
\begin{equation}
h_{\textnormal{3PN}}(f)={\cal A}_\textnormal{3PN}(f)\ e^{i(\psi_\textnormal{PP}+\psi_\textnormal{T})}
\end{equation}
is the standard TaylorF2 approximant of the GW signal in the frequency domain \cite{DIS00}. 
In this work we consider optimally oriented observers, such that the 3 PN amplitude reads:
\begin{align}
{\cal A}_\textnormal{3PN}(f)&={\cal A}f^{-7/6}
\sum_{k=0}^{6}\beta_{k}(m\pi f)^{k/3}\ ,\nonumber\\
&=\sqrt{\frac{5}{24}}\frac{{\cal M}^{5/6}}{\pi^{2/3}d}f^{-7/6}
\sum_{k=0}^{6}\beta_{k}(m\pi f)^{k/3}\ ,\label{ampl}
\end{align}
where ${\cal M}=m\nu^{3/5}$ is the chirp mass, $m=M_{\tn{BH}}+M_{\tn{NS}}$ and 
$\nu=M_{\tn{BH}}M_{\tn{NS}}/m^{2}$, and $d$ is the source distance. 
The coefficients $\beta_{k}$ are given by Eq.~(42) of \cite{DNT11}\footnote{We note that 
${\cal A}_\textnormal{3PN}$ contains some imaginary contributions; hereafter, when 
referring to the GW spectrum $f h(f)$, we shall mean the modulus $\vert f  h(f)\vert$.}. 
Finally, $\psi_\textnormal{PP}$ represents the point-particle contribution to the GW phase, 
currently known at the 3.5 PN order \cite{B06}, while $\psi_\textnormal{T}$ describes 
the effects of tidal interactions \cite{VFH11,BDF12}, and it is given by 
\begin{align}\label{tidalphase}
\psi_{\tn{T}}=&-\frac{117}{8\nu}\frac{\tilde\lambda}{m^5} x^{5/2}(1+2.5x-\pi x^{3/2}\nonumber\\
&+8.51x^{2}-3.92\pi x^{5/2})\ , 
\end{align} 
where $x=(m\pi f)^{2/3}$; the rescaled tidal deformability $\tilde\lambda$ is
related to the tidal deformability $\lambda$ (defined in Eq.~(\ref{deflambda}))
by
\begin{equation}\label{tidaldef}
\tilde\lambda =\frac{1+12q}{26}\lambda\ ,
\end{equation}
in the case  of BH-NS binaries, and by
\begin{equation}\label{tidaldef1}
\tilde\lambda =\frac{1}{26}\left[(1+12q)\lambda_{\tn{NS}_1}+\frac{q+12}{q}\lambda_{\tn{NS}_2}
\right]\ ,
\end{equation}
for NS-NS binaries (note that in this case, if $q=1$ $\tilde\lambda=\lambda$).
The function $\Theta(f,f_{\tn{cut}})$ appearing in the extended template
(\ref{hfit}) reproduces the sharp decrease in the amplitude corresponding to
tidal disruption; we choose it to have the form
\begin{equation}
\Theta(f,f_{\tn{cut}})=e^{-\alpha(f/f_\tn{cut}-1)}\ .\label{deftheta}
\end{equation}
We have also considered different forms of the cut-off function $\Theta$,
finding that such change does not significantly affect our results.  To compare
the approximate waveform given in (\ref{hfit}) with the waveforms produced by
numerical simulations of binary coalescence, $h_{\tn{NR}}$, it is useful to
compute the overlap of the two signals, given by
\begin{equation}
{\cal O}(\bar{h}_{\tn{PN}},h_\tn{NR})=\frac{(\bar{h}_{\tn{PN}}\vert h_\tn{NR})}{\sqrt{(\bar{h}_{\tn{PN}}
\vert \bar{h}_{\tn{PN}})}\sqrt{(\bar{h}_{\tn{NR}}\vert \bar{h}_{\tn{NR}})}}\ ,\label{defcalO}
\end{equation}
where (see Appendix \ref{app:Fisher})
\begin{equation}
(g\vert h)=2\int\frac{\tilde h(f)\tilde g^{\star}(f)+\tilde h^{\star}(f)\tilde g(f)}{S_{h}(f)}df\ ,
\end{equation}
$S_h(f)$ is the detector noise spectral density, and the integration is
performed in the frequency range $[f_{\tn{cut}},2f_{\tn{cut}}]$.  It is
worth remarking that although the gravitational waveform given by
Eq.~(\ref{hfit}) represents only a coarse approximation of the true signal
around the tidal disruption frequency, it is accurate enough for the study we
intend to carry on in this work.


For binary sources at cosmological distances, which are the main target of the
third generation detector ET, the GW signal has to be properly redshifted.  As
discussed in \cite{CF94,MR11}, the point particle phase $\psi_{\tn{PP}}$ and the
signal amplitude $\cal A_{\tn{3PN}}$ are invariant under the general
transformation $(f,{\cal M},d,t)\rightarrow (f/\xi,{\cal M}\xi,d\xi,t\xi)$,
where $\xi=1+z$. This means that using the point-particle approximation we can
only set constraints on the redshifted chirp-mass ${\cal M}_{z}={\cal M}(1+z)$
and on the luminosity distance $d_{\tn{L}}=d(1+z)$: if the proper distance $d$
is unknown, it is impossible to disentangle the mass parameter and the redshift.
However, the tidal phase $\psi_{\textnormal{T}}$ depends on the un-redshifted,
rest-frame mass components $M_{\textnormal{BH}},M_{\textnormal{NS}}$, and then
allows to break this degeneracy.

In this work we consider binary systems up to $z\simeq0.5$. The redshift $z$ is
given in terms of the distance $d$ by the relation
\begin{equation}
d=\frac{1}{H}\int_{0}^{z}\frac{dz'}{\sqrt{\Omega_{m}(1+z')^3+\Omega_\Lambda}}\ ,
\end{equation}
where $H=68$ km/Mpc/s is the Hubble constant, $\Omega_{m}=0.317$ is the total
matter density, and $\Omega_\Lambda=0.683$ is the dark energy density measured
by the Planck satellite \cite{Planck}.  In order to include the cosmological
redshift in the waveform, we (i) replace the quantities $(f,m,d)$ with the
redshifted quantities $(\bar f,\bar m,d_{\tn{L}})$ rescaled with $\xi=1+z$ as
discussed above (this does not change the functional form of the waveform); (ii)
modify the form of the tidal contribution and of the cut-off,
Eqns.~(\ref{tidalphase}) and (\ref{deftheta}), as follows: \begin{align}
  \psi_{\tn{T}}(\bar f)=&-\frac{117}{8\nu}\frac{(1+z)^5\tilde\lambda}{\bar m^5}
  x^{5/2}(1+2.5x-\pi x^{3/2}\nonumber\\ &+8.51x^{2}-3.92\pi x^{5/2})
\label{tidalphasez}\\ \Theta(\bar f,f_{\tn{cut}})=&e^{-\alpha(\bar
f(1+z)/f_\tn{cut}-1)}\,.\label{defthetaz} \end{align} 

Once the GW signal from a NS-NS or NS-BH inspiral has been detected, it will be
possible to extract the values of the binary parameters comparing the data with
the GW template, through a matched filtering technique. Assuming Gaussian noise
and sufficiently high signal-to-noise ratio, the parameter variance can be
estimated using the Fisher matrix approach (see for example
\cite{CF94,PW95,FH08,HLLR10,DNV12}, and Appendix~\ref{app:Fisher} of this
paper), which we here discuss.

In the next Section we shall restrict to the case of GW-signals emitted by a NS-BH
coalescing binary which exhibits the feature associated to NS disruption,
i.e. $f_{\tn{cut}}$.  In this case, the GW template defined in Eq.~(\ref{hfit})
depends on the set of parameters
\begin{equation}\label{param3}
\boldsymbol\theta=(\ln{\cal A},t_{c},\phi_{c},\ln{\cal M},\ln\nu,\tilde\lambda,f_{\textnormal{cut}})\ ,
\end{equation}
where $t_c$ and $\phi_c$ are the time and the phase at the coalescence.  We do
not include the cosmological redshift $z$ among the set of parameters
(\ref{param3}), since we assume that $z$ is known a priori by coincidence
measurements in the electromagnetic band. Therefore, our computation will
eventually yield the uncertainties on the {\it un-redshifted} parameters.

The parametrization (\ref{param3}) leads to a $7\times 7$ covariance matrix
(see Appendix~\ref{app:Fisher}), whose diagonal elements represent the standard
deviation or the parameters (\ref{param3}). We remark that $\ln {\cal A}$ is
uncorrelated with the other parameters \cite{PW95}; therefore, we will restrict
our analysis to the remaining $6$ parameters. We also remark that once
$\tilde\lambda$, ${\cal M}$ and $\nu$ are estimated, the tidal deformability
$\lambda$ can be computed from Eq.~(\ref{tidaldef}) or (\ref{tidaldef1}).

Since, as we shall show in Sec.~\ref{sec:model}, the cut-off frequency
$f_{\tn{cut}}$ will mostly be accessible to third generation interferometers, in
the following analysis we will use the noise spectral density of the Einstein
Telescope described by the fit
\begin{align}
  \frac{S_h}{S_{0}}=x^{-4.1}&+186x^{-0.69}+233[1+31x-65x^{2}+52x^{3} \ +\nonumber\\
  &-42x^4+10x^{5}+12x^{6}]/[1+14x-37x^{2}\ + \nonumber\\
  &+19x^{3}+27x^{4}]
\label{ETLfit}
\end{align}
with $\bar f\geq f_{s}=10$ Hz, $S_{0}=10^{-52}$ Hz$^{-1}$ and $x=\bar f/f_{0}$, being
$f_{0}=200$ Hz a scaling frequency \cite{SC09}.

\section{Data-analysis strategies}\label{sec:strategies}
In this Section we discuss how the cut-off frequency $f_{\tn{cut}}$, identified
in a detected gravitational wave signal emitted by a BH-NS binary, can be used
to extract information on the NS EoS.  As discussed in
Section~\ref{sec:template}, the gravitational waveform depends on two quantities
which carry the imprint of the NS EoS: the deformability $\lambda$ and - if
tidal disruption occurs - the cut-off frequency $f_{\tn{cut}}$. These quantities
can both be related to the NS compactness ${\cal C}$, using the
$f_{\tn{cut}}({\cal C})$ fit (\ref{fcutC}) found in \cite{KST10}, and the
$\lambda({\cal C})$ fit found in \cite{MCFGP13}:
\begin{equation}\label{ILQ} 
{\cal C}=0.371-0.0391\ln \left[\frac{\lambda}{M_{\textnormal{NS}}^{5}}\right]
+0.001056\left[\ln\left(\frac{\lambda}{M_{\textnormal{NS}}^{5}}\right)\right]^2\ ,
\end{equation} where $\lambda(\tilde\lambda)$ is given by
Eq.~(\ref{tidaldef}).  We remind that Eq.~(\ref{ILQ}) is found to
reproduce the values of the star compactness with an accuracy greater
3\%, for a large class of EoS. In the following, we shall denote by
${\cal C}_\lambda$ the NS compactness obtained from the fit (\ref{ILQ}),
and by ${\cal C}_{\tn{cut}}$ the NS compactness obtained from the fit
(\ref{fcutC}).

We propose and compare two data-analysis strategies to extract information on
the neutron star equation of state, which employ these two fits in a different
way.

With the first strategy, we assume that the unknown parameters are
\begin{equation}\label{param4}
\boldsymbol\theta=(t_{c},\phi_{c},\ln{\cal
M},\ln\nu,\tilde\lambda,f_{\textnormal{cut}})\ ,
\end{equation}
and estimate the corresponding errors $\sigma_i, i=1,6$, 
with a $6\times6$ covariance matrix.  We then write the relation (\ref{fcutC})
between $f_\tn{cut}$ and the NS compactness, in the form
\begin{equation}
\ln(f_\textnormal{cut}m)=(a\pm \sigma_{a})\ln {\cal C}_{\tn{cut}}+(b\pm \sigma_{b})\ ,
\end{equation}
where $a,\sigma_a,b,\sigma_b$ are given in eq. (\ref{fcutC}).  Assuming the
errors to be uncorrelated, we compute the variance on ${\cal C}_{\tn{cut}}$ as:
\begin{equation}\label{Cfcut}
\sigma^{2}_{{\cal C}_{\tn{cut}}}=\sum_{p_{i}}\left(\frac{\partial {\cal C}_{\tn{cut}}}{\partial p_{i}}
\right)^{2}\sigma^{2}_{p_{i}}\ ,
\end{equation}
where $p_{i}=\{a,b,m,f_{\tn{cut}}\}$
\footnote{
The error on the total mass $m$ is easily computed from 
the chirp mass ${\cal M}$ and the symmetric mass ratio $\nu$ as 
\begin{equation}\label{sm}
  \sigma^{2}_{m}=\left(\frac{\partial m}{\partial {\cal M}}\right)^2\sigma^2_{\cal M}+
  \left(\frac{\partial m}{\partial \nu}\right)^2\sigma^2_{\nu}+2
  \frac{\partial m}{\partial {\cal M}}\frac{\partial m}{\partial {\cal \nu}}
  cov({\cal M},\nu)\ ,
\end{equation}
where $cov({\cal M},\nu)$ is the ${\cal M}$-$\nu$ element of the covariant
matrix for the parameters (\ref{param4}).}.  Eq.~(\ref{Cfcut}) provides the
error on the NS compactness derived from the frequency cut-off only.

Using the fit (\ref{ILQ}) we then compute the error on the NS compactness
directly from the tidal deformability $\lambda$, as
\begin{equation}
\sigma_{{\cal C}_{\lambda}}^{2}=\sigma^{2}_{\tn{fit}}+\left(\frac{\partial {\cal C}_{\lambda}}{\partial \ln\tilde\lambda}\right)^{2}
\sigma^{2}_{\ln\tilde\lambda}+\left(\frac{\partial {\cal C}_{\lambda}}{\partial M_{\tn{NS}}}\right)^{2}
\sigma^{2}_{M_{\tn{NS}}}\label{properr}
\end{equation}
where we set $\sigma_{\tn{fit}}=0.03\,{\cal C}$. The latter has been estimated
in \cite{MCFGP13} and it corresponds to the largest relative discrepancy between
the value of $\lambda$ obtained from the fit and the value computed solving the
equations of stellar perturbations, for a set of EoS covering a large range of
stiffness. In addition, since the two masses are comparable (in this paper we
only consider low values of the mass ratio) and the total error is dominated by
$\sigma_{\tn{fit}}$, we assume $\sigma_{M_{\tn{NS}}}$ to be of the same order of
$\sigma_{m}$. The value of the NS compactness ${\cal C}$ is obtained combining
the information given by the two fits (\ref{Cfcut}) and (\ref{ILQ}), and is the
weighted mean
\begin{equation}\label{meanC}
{\cal C}=\frac{{\cal C}_{\tn{cut}}/\sigma^{2}_{{\cal C}_{\tn{cut}}}+{\cal
C}_{\lambda}/\sigma^{2}_{{\cal C}_\lambda}}{\sigma^{-2}_{{\cal C}_{\tn{cut}}}
+\sigma^{-2}_{{\cal C}_\lambda}}\ ,
\end{equation}
with variance
\begin{equation}
\sigma^2_{{\cal C}}={\frac{1}{\sigma_{{\cal C}_{\tn{cut}}}^{-2}+\sigma_{{\cal C}_\lambda}^{-2}}}\ .
\end{equation}

The second strategy consists in expressing the information on the neutron star
internal composition in terms of one single parameter, the rescaled tidal
deformability $\tilde\lambda$.  We combine the fits (\ref{fcutC}) and
(\ref{ILQ}) (using Eq.~(\ref{tidaldef})) to express $f_{\tn{cut}}$ in terms of
$\tilde\lambda$, and substitute this expression in the waveform template given
by eq.~(\ref{hfit}).  We then compute the $5\times5$ covariance matrix, for the
set of variables $\boldsymbol\theta=(t_{c},\phi_{c},\ln{\cal
  M},\ln\nu,\tilde\lambda)$.  The value of the NS compactness ${\cal C}$ and its
variance are then obtained from the fit (\ref{ILQ}) and
Eq.~(\ref{properr}). This procedure reduces the data-analysis computational cost
with respect to the previous strategy, since it does not include the parameter
$f_{\tn{cut}}$.

\subsection{The binary models}\label{sec:binary_model}
In order to be fully consistent with the fit (\ref{fcutC}) derived in
\cite{KST10}, in this Section we shall consider the same binary models used in
\cite{KST10}, i.e.  BH-NS binaries with mass ratio $q=2$, with the neutron star
modeled using a set of piecewise polytropes \cite{RMSUCF09}.  In particular,
the core is described by a polytropic $p=K\rho^\Gamma$ with $\Gamma=3$, and such
that $p$ is equal to a given value $p_1$ at density $\rho_{1}=5.0119\times
10^{14}$g cm$^{-3}$; the crust is described by a polytropic with
$\Gamma=1.3569$, such that $p=1.5689\times 10^{31}$dyn cm$^{-2}$ at density
$\rho=10^{13}$g cm$^{-3}$.  The overall EoS depends on the value of the
parameter $p_1$. In \cite{KST10} four EoS have been considered, corresponding to
four values of $p_1$, and named (from the stiffest to the softest) \texttt{2H},
\texttt{H}, \texttt{HB} and \texttt{B}.

\begin{figure}[!h]
\centering
\includegraphics[width=8.3cm]{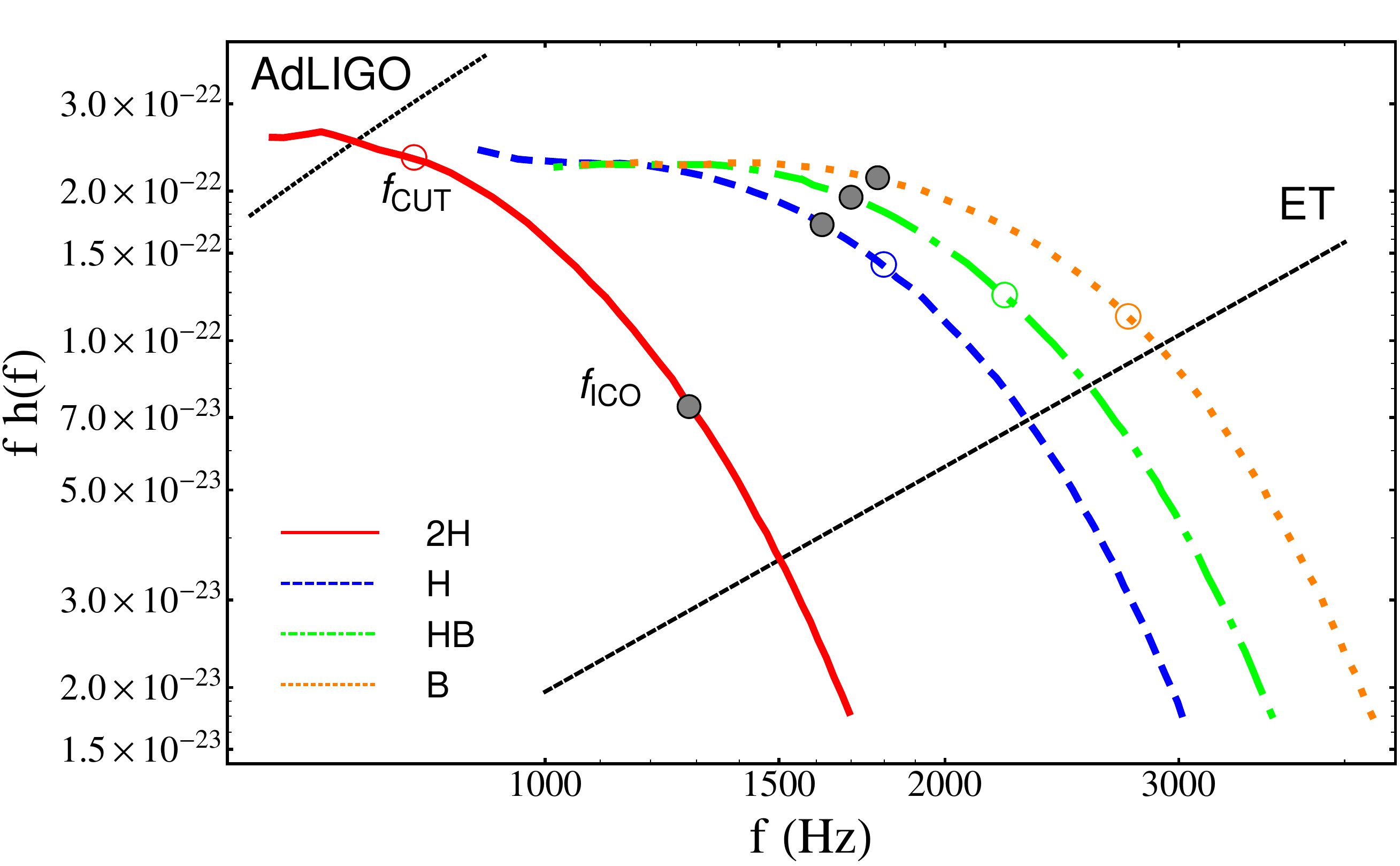}
\caption{Gravitational wave spectra $f h(f)$ extracted from \cite{KST10}, for 
$q=2$ and $M_{\textnormal{NS}}=1.2M_{\odot}$ at a distance $d=100$ Mpc. 
Empty and grey circles refer to the cut-off and the ICO frequencies. 
We also show the noise spectral density of 
Advanced LIGO and the Einstein Telescope.}
\label{plot_fit2}
\end{figure}

In Fig.~\ref{plot_fit2} we show the gravitational wave spectra $f h(f)$
extracted from figure 9 of \cite{KST10}; the spectra refer to BH-NS systems with
mass ratio $q=2$ and $M_{\textnormal{NS}}=1.2M_{\odot}$ located at the distance
$d=100$ Mpc; in addition we plot the noise spectral densities of Advanced
LIGO \cite{LIGOVirgo} and of the Einstein Telescope (ET) \cite{ET}, the
values of the cut-off frequency (empty circles) derived from Eq.~(\ref{fcutC}),
and those of the Innermost Circular Orbit (ICO - grey circles), computed by
minimizing the total Post-Newtonian binding energy of the two body system as in
\cite{B06,VF10}.  From Fig.~\ref{plot_fit2} we can immediately draw the
following considerations: (i) the NS is tidally disrupted before the ICO only
for the stiffest EoS, \texttt{2H}; (ii) for the considered systems, third
generation detectors like ET are needed to unambiguously detect $f_{\tn{cut}}$.
We have also considered binaries with $M_{\textnormal{NS}}=1.35M_{\odot}$,
finding that the same conclusions apply. Moreover, increasing the neutron star
mass, the compactness increases, leading to higher cut-off frequencies.

Therefore, in applying the proposed data analysis strategies we shall only
consider NSs modeled by the $\texttt{2H}$ EoS (defined by $p_{1}=10^{13.95}$
g/cm$^3$), with $M_{\tn{NS}}=1.20\,M_\odot, R_{\tn{NS}}=15.10$ km and
$M_{\tn{NS}}=1.35\,M_\odot, R_{\tn{NS}}=15.20$ km.  From the fit (\ref{fcutC})
we find $f_{\tn{cut}}=796.67$ Hz and $f_{\tn{cut}}=1085.51$ Hz,
respectively. Using the approach of \cite{H08} we have computed the Love number
for these configurations, finding $k_2=0.1452$ and $k_2=0.1342$, respectively.
As in \cite{KST10}, we have considered non-spinning BHs with mass ratio
$q=M_{\tn{BH}}/M_{\tn{NS}}=2$. As discussed in Section \ref{sec:cutoff}, the
analysis of this work can easily be extended to more general configurations,
once fully relativistic simulations will provide a fit analogous to
Eq.~(\ref{fcutC}).

The coefficient $\alpha$ in Eq.~(\ref{deftheta}) has been chosen to minimize the
discrepancy between the numerical spectra given in \cite{KST10} and the
analytical spectra given by eq. (\ref{hfit}).  This value is $\alpha=1.55$.  As
a further check, we have compared the waveform given in (\ref{hfit}) with those
obtained from the numerical simulations of \cite{KST10}, computing the overlap
integrals (\ref{defcalO}) in the frequency range $[f_{\tn{cut}},f_{\tn{end}}]$
(where $f_{\tn{end}}$ is the highest value of frequency in the numerical
data). We find ${\cal O}(\bar{h}_{\tn{PN}}, h_\tn{NR})=0.9991$ for
$M_{\textnormal{NS}}=1.2M_{\odot}$ and ${\cal O}(\bar{h}_{\tn{PN}},
h_\tn{NR})=0.9997$ for $M_{\textnormal{NS}}=1.35M_{\odot}$. These values are
compatible with the standard accuracy threshold used for GW detection, i.e.
$1-{\cal O}<0.005$ \cite{LOB08} (assuming that the same threshold will hold for
the ET data analysis). We note that, if we compute the overlap between the
waveform containing only the inspiral part of the signal $h_{\tn{3PN}}$, and the
numerical one, we obtain ${\cal O}(h_\tn{3PN},h_\tn{NR})=0.976333$ and ${\cal
  O}(h_\tn{3PN},h_\tn{NR})=0.973471$, respectively.  Choosing functional forms
for the cut-off function $\Theta$ compatible with the numerical data, but
different from eq.~(\ref{deftheta}), the overlap integrals do not change
significantly.

\subsection{Results}\label{sec:results}

In Table~\ref{configs} we summarize the parameters of the binary configurations
we consider, i.e.  non-spinning BH-NS binaries with mass ratio $q=2$, EOS
\texttt{2H}, $M_{\tn{NS}}=(1.2,1.35)M_{\odot}$; we assume they are located at
fixed luminosity distances $d_{\tn{L}}=100/500/1000/2000$ Mpc, and identify the
system with NS mass $\tn{M}_\tn{NS}$ at distance $d_{\tn{L}}$, with the label
\texttt{2H}\_\texttt{d}$_\tn{L}$\_\texttt{M}$_{\textnormal{NS}}$ (first column).
In column 2 and 3 we show the NS compactness ${\cal C}$ and the rescaled tidal
deformability $\tilde\lambda$ given by Eq.~(\ref{tidaldef}), respectively.
Since, as discussed in Sec.~\ref{sec:template}, we take into account the effect
of cosmological redshift, the values of the redshift $z$ and of the redshifted
cut-off frequencies $\bar{f}_{\tn{cut}}=f_{\tn{cut}}/(1+z)$ are also shown in
Table~\ref{configs}; in the last column we show the signal-to noise ratio
$\rho$, assuming the detector is ET.
\begin{table}[h!]
\centering
\begin{tabular}{cccccccc}
\hline
\hline
\texttt{model} & ${\cal C}$ & $\tilde\lambda$ (km$^5$)  &$z$ & $\bar{f}_{\tn{cut}}$ (Hz) & $\rho$ \\
\hline
\texttt{2H\_100\_120} &  0.117 & 7.31$\cdot 10^{4}$    &0.023 &  779 &563  \\
\texttt{2H\_500\_120}  & &            &0.117 &  713 &121\\
\texttt{2H\_1000\_120} & &            &0.240 &  642 &66\\
\texttt{2H\_2000\_120} & &            &0.519 &  524 &39\\
\hline                                     
\texttt{2H\_100\_135} & 0.131 & 6.98$\cdot 10^{4}$    &0.023 &  1061 &619\\
\texttt{2H\_500\_135}  & &            &0.117 &   972 &133 \\
\texttt{2H\_1000\_135} & &            &0.240 &   875 & 72 \\
\texttt{2H\_2000\_135} & &            &0.519 &   715 & 42 \\
\hline
\hline
\end{tabular}
\caption{The NS compactness ${\cal C}$,  rescaled tidal deformability
$\tilde\lambda$, cosmological redshift,  redshifted cut-off 
frequencies and signal-to noise ratio (assuming the detector is ET)
are shown for the binary configurations considered in this section. }\label{configs}
\end{table}


In Table~\ref{errors} we show the errors obtained, applying the Fisher matrix
approach, on the chirp mass $\cal M$, the symmetric mass ratio $\nu$, the
rescaled tidal deformability $\tilde\lambda$, and the cut-off frequency
$f_{\tn{cut}}$, for the binary configurations shown in Table~\ref{configs}.  In
the last column we show the error on the total mass of the system, $m$. The
integrations on frequency (see Appendix \ref{app:Fisher}) have been performed in
the range $[10\,{\rm Hz},2\,f_{\tn{cut}}]$.  Table~\ref{errors} shows that the
relative errors on the parameters increase of about a factor ten going from
$100$ Mpc to $2$ Gpc.
\begin{table}[ht!]
\centering
\begin{tabular}{cccccc}
\hline
\hline
\texttt{model} & $\sigma_{\ln {\cal M}}$(\%) & $\sigma_{\ln \nu}$(\%) &  
$\sigma_{\ln\tilde\lambda }$ (\%) & $\sigma_{\ln f_{\tn{cut}}}$(\%) & $\sigma_{\ln m}$(\%)\\
\hline
\texttt{2H\_100\_120}    & 1.0$\cdot 10^{-4}$ & 2.0$\cdot 10^{-2}$ &  1.4  & 3.6  & 1.2 $\cdot 10^{-2}$ \\
\texttt{2H\_500\_120}    & 5.7$\cdot 10^{-4}$ & 9.7$\cdot 10^{-2}$ &  6.0  & 14   & 5.8 $\cdot 10^{-2}$ \\
\texttt{2H\_1000\_120}  & 1.3$\cdot 10^{-3}$ & 1.9$\cdot 10^{-1}$ &  10   & 22   & 1.2 $\cdot 10^{-1}$  \\
\texttt{2H\_2000\_120}  & 3.2$\cdot 10^{-3}$ & 3.9$\cdot 10^{-1}$ &  15   & 27   & 2.3 $\cdot 10^{-1}$  \\
\hline                                     
\texttt{2H\_100\_135}    & 1.1$\cdot 10^{-4}$ & 1.7$\cdot 10^{-2}$ &  1.6   & 6.6 & 1.0 $\cdot 10^{-2}$ \\
\texttt{2H\_500\_135}    & 6.1$\cdot 10^{-4}$ & 8.6$\cdot 10^{-2}$ &  6.7   & 26  & 5.1 $\cdot 10^{-2}$ \\
\texttt{2H\_1000\_135}  & 1.4$\cdot 10^{-3}$ & 1.7$\cdot 10^{-1}$ &  11    & 40  & 1.0 $\cdot 10^{-1}$  \\
\texttt{2H\_2000\_135}  & 3.4$\cdot 10^{-3}$ & 3.4$\cdot 10^{-1}$ &  17    & 49  & 2.0 $\cdot 10^{-1}$  \\
\hline
\hline
\end{tabular}
\caption{Percentage errors on the chirp mass, symmetric mass ratio,
  rescaled tidal deformability, cut-off frequency and total mass for the
  configurations listed in the first column.}\label{errors} \end{table}
It should be stressed that the analysis shows that $f_{\tn{cut}}$ is
uncorrelated with the other parameters. Therefore, for instance, the value of
$\sigma_{\ln\tilde\lambda }$ is the same we would obtain using the waveform
template given by Eq. (\ref{hfit}) without $f_{\tn{cut}}$.

The results obtained using the first strategy to estimate the NS compactness and
the corresponding error, are presented in Table~\ref{strat1}, where we show the
relative percentage errors on: (i) the compactness ${\cal C}_{\tn{cut}}$ derived
from the frequency cut-off and the analytic fit (\ref{fcutC}), (ii) the
compactness ${\cal C}_\lambda$ estimated by means of the universal relation
(\ref{ILQ}), (iii) the weighted compactness ${\cal C}$ defined in
Eq.~(\ref{meanC}). We do not show explicitly the value of the compactness
obtained from this approach, because it coincides (with a discrepancy smaller
than $1\%$) with the ``true'' value shown in Table~\ref{configs}.
\begin{table}[h!]
\centering
\begin{tabular}{cccccc}
\hline
\hline
\texttt{model}
& $\sigma_{\ln{\cal C}_{\tn{cut}}}$(\%) & $\sigma_{\ln{\cal C}_\lambda}(\%)$  
& $\sigma_{\ln{\cal C}}$(\%)\\
\hline
\texttt{2H\_100\_120}  
& 8.8   & 3.0   & 2.8  \\
\texttt{2H\_500\_120}   
& 9.5   & 3.2   & 3.0  \\
\texttt{2H\_1000\_120} 
& 10 & 3.5   & 3.3  \\
\texttt{2H\_2000\_120} 
& 11 & 4.1   & 3.9  \\
\hline                             
\texttt{2H\_100\_135}   
& 8.7   & 3.0   & 2.8  \\
\texttt{2H\_500\_120}   
 & 11 & 3.2   & 3.1  \\
\texttt{2H\_1000\_120} 
 & 14 & 3.6   & 3.5  \\
\texttt{2H\_2000\_120} 
 & 16 & 4.1   & 4.0  \\
\hline
\hline
\end{tabular}
\caption{Relative percentage errors on the compactness ${\cal C}$
derived from the frequency cut-off fit (\ref{fcutC}), from the
universal relation (\ref{ILQ}),
and from the weighted mean (\ref{meanC}).}\label{strat1} \end{table}

\begin{figure}[h!]
\centering
\includegraphics[width=8cm]{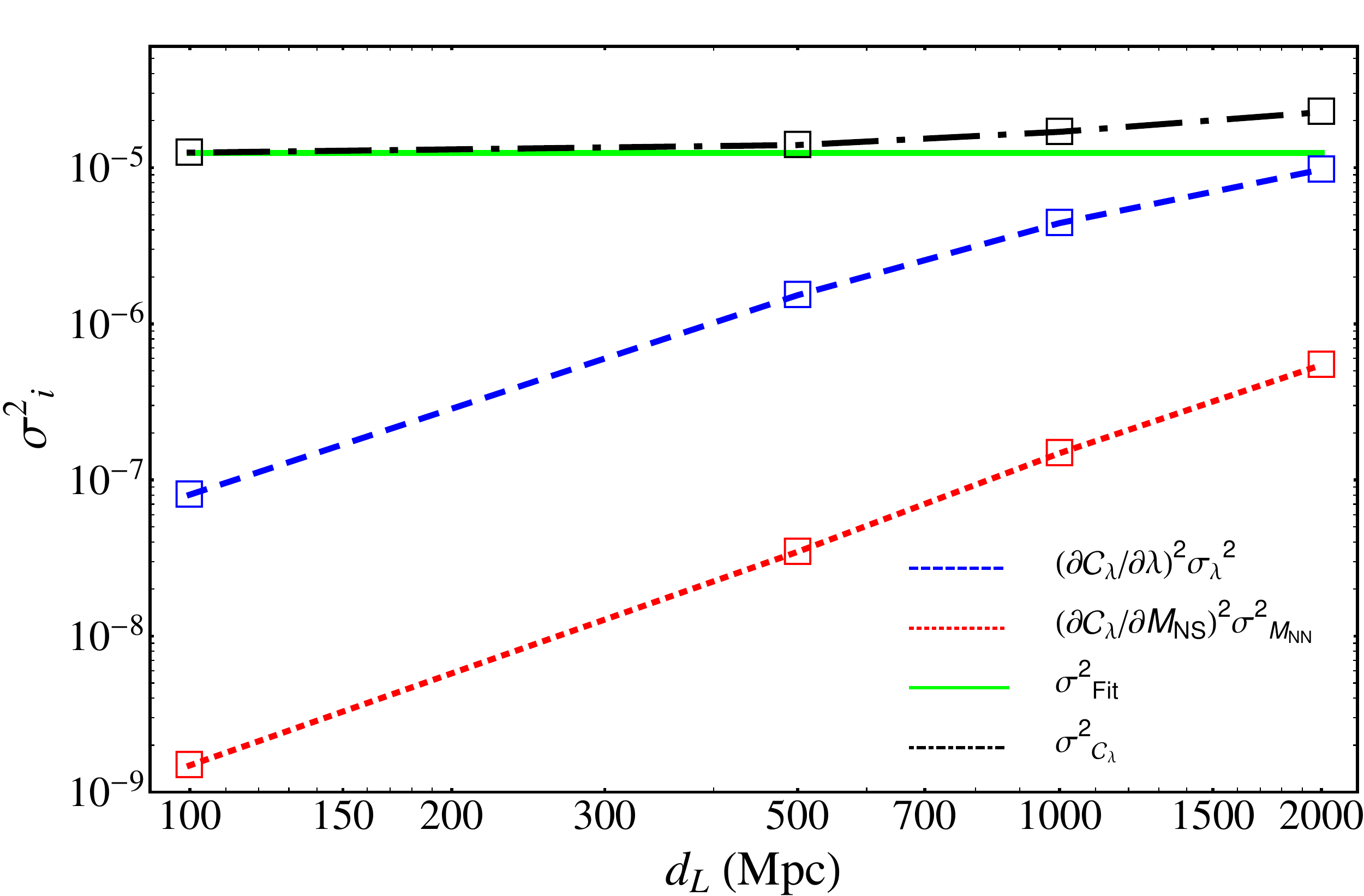}
\caption{We show the three terms appearing in Eq.~(\ref{properr}), contributing to  
the relative error on $\sigma_{{\cal C}_{\lambda}}$, 
as functions of the luminosity distance $d_{\tn{L}}$, for 
$M_{\tn{NS}}=1.2M_{\odot}$. The dot-dashed black curve 
represents the total error $\sigma^2_{{\cal C}_{\lambda}}$.}
\label{plot_errors}
\end{figure}

From Table~\ref{strat1} we see that the relative percentage error in the NS
compactness is of the order of $3\%-4\%$, with a very mild increase with the
source distance. This weak dependence on $d_{\tn{L}}$ is due to the functional
form of $\sigma_{{\cal C}_{\lambda}}$ which is dominated by the fit error
$\sigma_{\tn{fit}}$. Indeed, the latter is found to be always greater than the
other two terms of Eq.~(\ref{properr}) up to $d_{\tn{L}}=1$ Gpc, when $(\partial
{\cal C}_{\lambda}/\partial \ln\tilde\lambda)^{2} \sigma^{2}_{\ln\tilde\lambda}$
starts to be of the same order of magnitude of $\sigma_{\tn{fit}}^2$
\footnote{The error on the NS mass is always an order of magnitude smaller than
  the other two terms, for all the considered configurations.}. The behavior of
these quantities, as a function of the luminosity distance (for
$M_{\tn{NS}}=1.2M_{\odot}$), is shown in Fig.~\ref{plot_errors}.
We also see that the error $\sigma_{{\cal C}_{\lambda}}$ is reduced when
$f_{\tn{cut}}$ is included in the analysis, but this reduction is
marginal. Thus, if our goal is to estimate the stellar compactness, the use of
$f_{\tn{cut}}$ in the data analysis does not introduce any significant
improvement.

Let us now see whether using the second strategy is more convenient.  The
results are shown in Table~\ref{strat2}, where we tabulate the relative
percentage errors on the chirp mass ${\cal M}$, the symmetric mass ratio $\nu$,
the rescaled tidal deformability and the compactness obtained from the fit
(\ref{ILQ}).  As in the previous case, we do not explicitly show the value of
the compactness because it coincides with the value shown in
Table~\ref{configs}.
\begin{table}[ht!]
\centering
\begin{tabular}{cccccc}
\hline
\hline
\texttt{model} & $\sigma_{\ln {\cal M}}$(\%) & $\sigma_{\ln \nu}$(\%) & 
$\sigma_{\ln\tilde\lambda }$ (\%) & $\sigma_{\ln {\cal C}}$(\%)\\
\hline
\texttt{2H\_100\_120}    & 1.0$\cdot 10^{-4}$ & 1.9$\cdot 10^{-2}$  & 1.3   & 3.0 \\
\texttt{2H\_500\_120}    & 5.6$\cdot 10^{-4}$ & 9.4$\cdot 10^{-2}$  & 5.3   & 3.1 \\
\texttt{2H\_1000\_120}  & 1.2$\cdot 10^{-3}$ & 1.8$\cdot 10^{-1}$  & 8.2   & 3.3 \\
\texttt{2H\_2000\_120}  & 3.1$\cdot 10^{-3}$ & 3.5$\cdot 10^{-1}$  & 10 & 3.6 \\
\hline                                     
\texttt{2H\_100\_135}    & 1.1$\cdot 10^{-4}$ & 1.7$\cdot 10^{-2}$ & 1.5   & 3.0 \\
\texttt{2H\_500\_135}    & 6.0$\cdot 10^{-4}$ & 8.3$\cdot 10^{-2}$ & 6.1   & 3.2 \\
\texttt{2H\_1000\_135}  & 1.3$\cdot 10^{-3}$ & 1.6$\cdot 10^{-1}$ & 9.5   & 3.4 \\
\texttt{2H\_2000\_135}  & 3.3$\cdot 10^{-3}$ & 3.1$\cdot 10^{-1}$ & 12 & 3.6 \\
\hline
\hline
\end{tabular}
\caption{Relative percentage errors on the chirp mass, 
  symmetric mass ratio, rescaled tidal deformability, and compactness 
  computed the second data-analysis strategy proposed in this Section.}\label{strat2}
\end{table}
We can see that the second strategy yields similar errors for the NS
compactness, of the order of $3\%-4\%$.  Conversely, if we compare the relative
percentage error on $\tilde\lambda$ computed with this strategy (column 4 of
Table~\ref{strat2}), with the same given in column 4 of Table~\ref{errors}, we
see that the error is reduced when the second strategy is applied, and the
reduction is up to $\sim 30\%$ for more distant sources.  It may also be noted
that, unlike ${\cal{C}}$, $\sigma_{\tilde\lambda}$ varies from $\sim 1\%$ to
$\sim 10\%$.

At this point we may ask whether $\cal{C}$ is the best parameter to be used to
gain information on the NS equation of state. This problem will be discussed in
detail in the next Section.

\section{Choosing the most suitable quantity to constrain the NS EoS}\label{sec:lambdaC}
In this section we compare the relative error on the two parameters which allow
to constrain the equation of state of matter in the NS interior, i.e. $\lambda$
(or equivalently $\tilde\lambda$) and ${\cal C}$.  Tidal effects in NS-NS
coalescing binaries may be revealed by second generation detectors AdvLIGO/Virgo
within a distance of $100$ Mpc, provided the signal-to noise ratio is larger
than 16 \cite{DNV12}; therefore,we extend the analysis of the previous section,
based on the Fisher matrix approach, to NS-NS and BH-NS configurations, choosing
a larger set of EoS to model the NS interior, and considering the sensitivity
curves of both AdvLIGO/Virgo and of the Einstein Telescope.  We do not include
$f_{\tn{cut}}$ in this analysis, because its role has already been discussed in
Section~\ref{sec:results}.  We use $7$ different EoS, which are described in
terms of piecewise polytropes \cite{RMSUCF09}, with three pieces for the NS
inner core and one piece for the crust. We select: (i) two EoS derived using
variational-method approaches, \texttt{APR4} and \texttt{WFF1}
\cite{APR4},\cite{WFF1}; (ii) two based on the relativistic
Brueckner-Hartree-Fock approach, \texttt{MPA1} and \texttt{ENG}
\cite{MPA1},\cite{ENG}; (iii) one based on the potential-method \texttt{SLy4}
\cite{SLy4}; (iv) one derived within the relativistic mean-field approach
\texttt{MS1} \cite{MS1}; (v) one relativistic mean-field theory EoS which
includes hyperons \texttt{H4} \cite{H4}.  It is useful to remind that stiffer
EoS correspond to more deformable stars, larger values of $\tilde\lambda$ and
smaller values of ${\cal C}$.  Ordering our EoS from the softest to the
stiffest, we find \texttt{WFF1, APR4, SLy4, ENG, MPA1, H4, MS1}.

\begin{figure*}[htp] \centering \includegraphics[width=7.8cm]{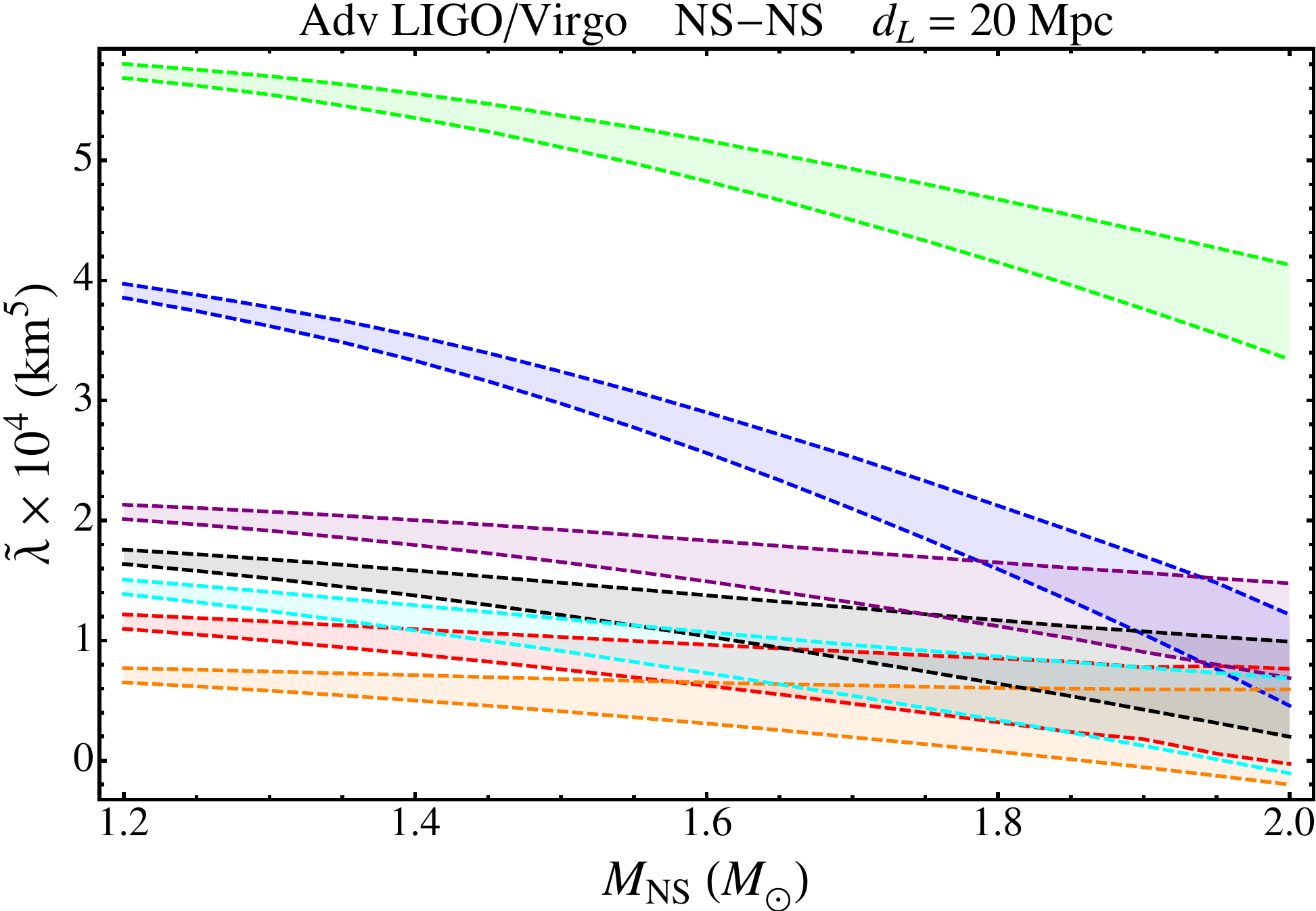}
\includegraphics[width=8.05cm]{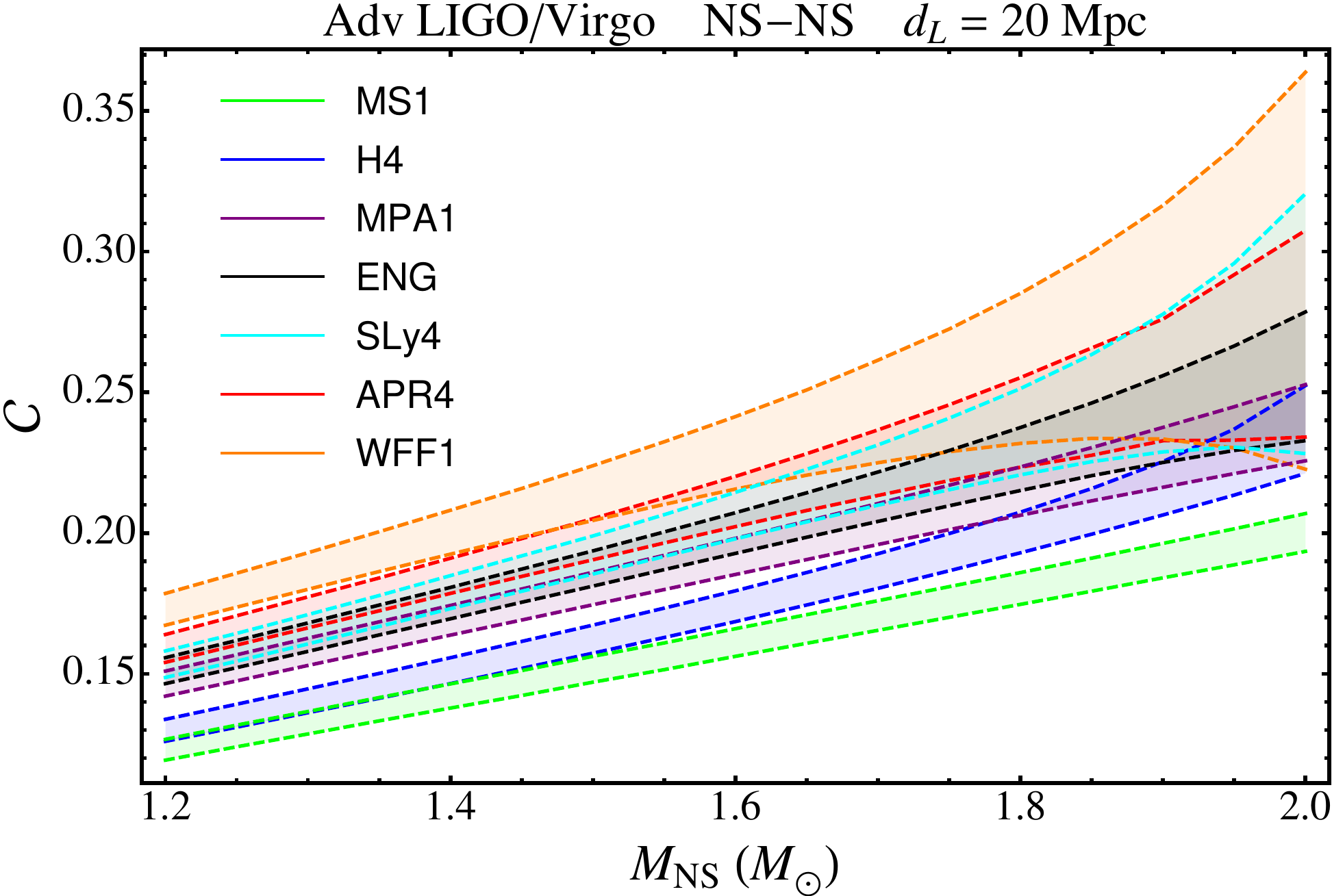}\\ \vspace{0.5cm}
\includegraphics[width=7.9cm]{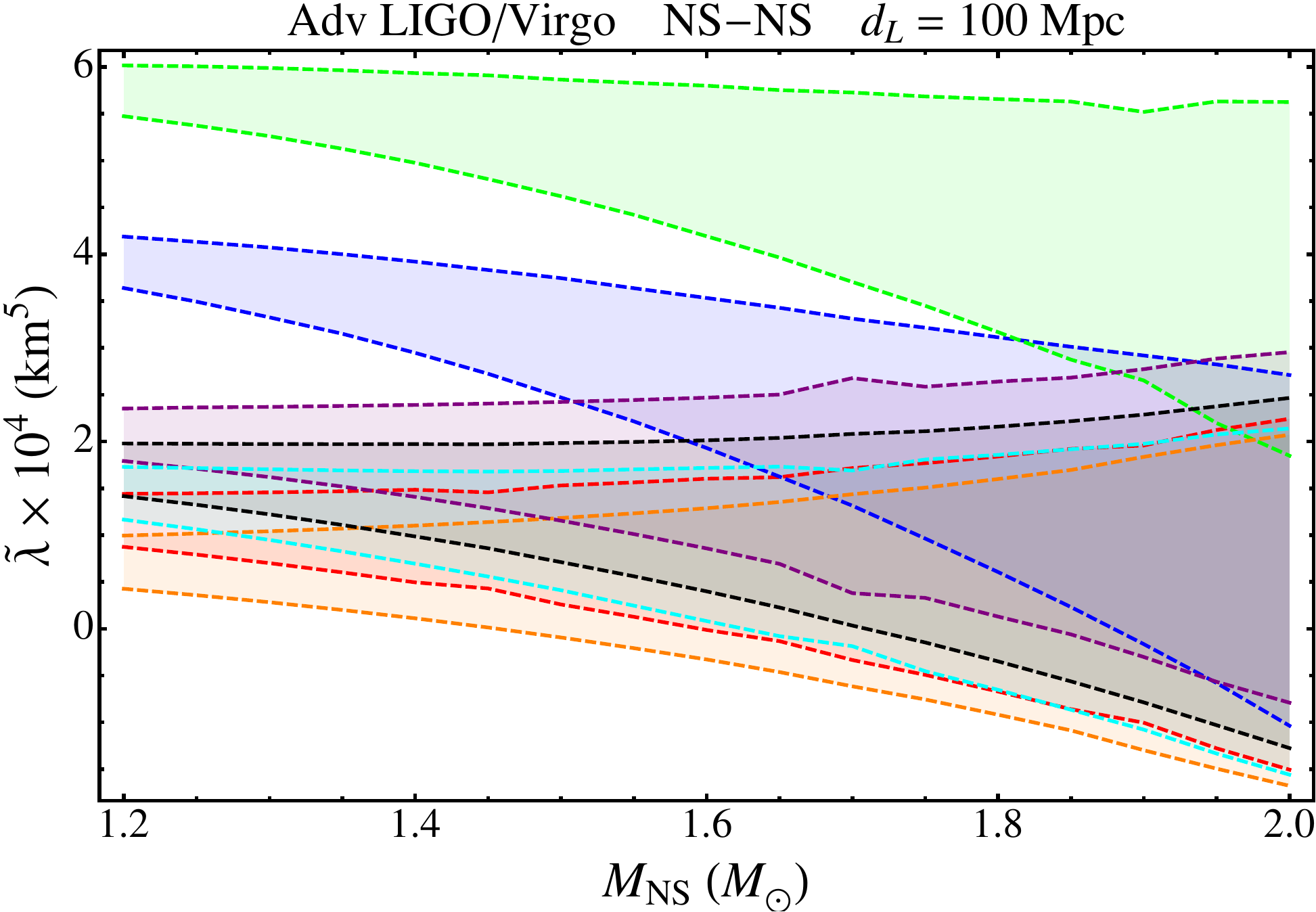}
\includegraphics[width=8.cm]{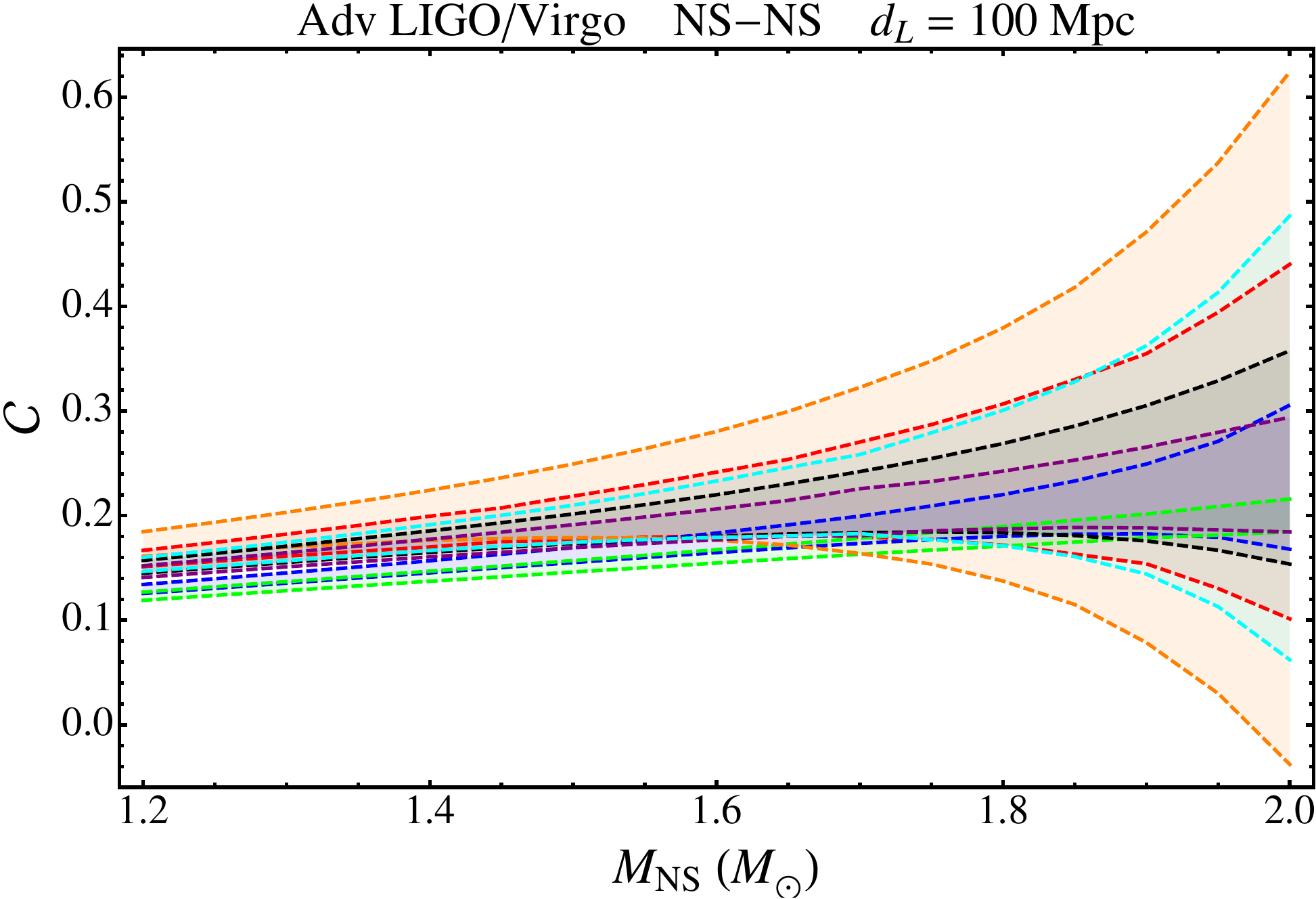}\\ \vspace{0.5cm}
\includegraphics[width=8.1cm]{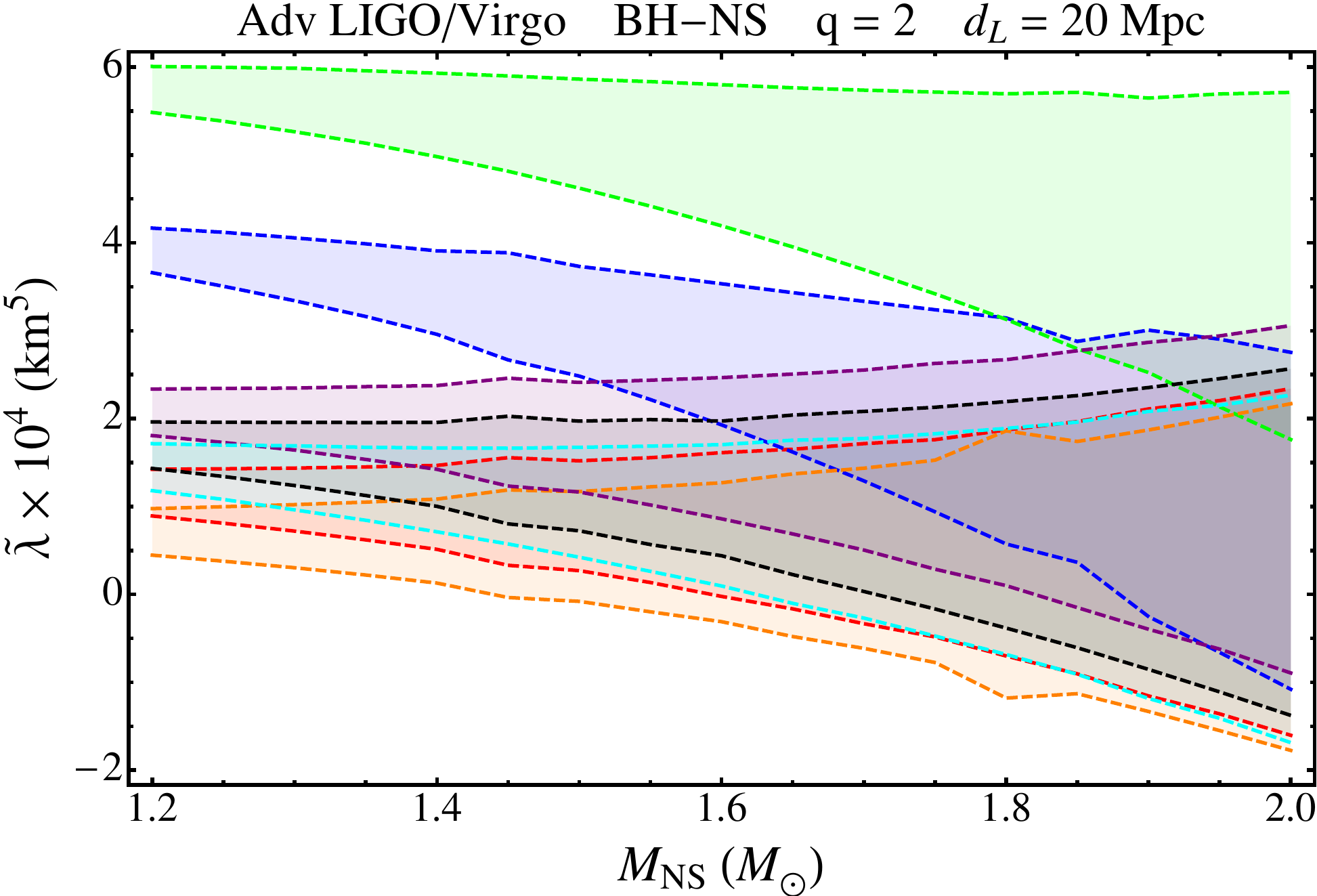}
\includegraphics[width=8.05cm]{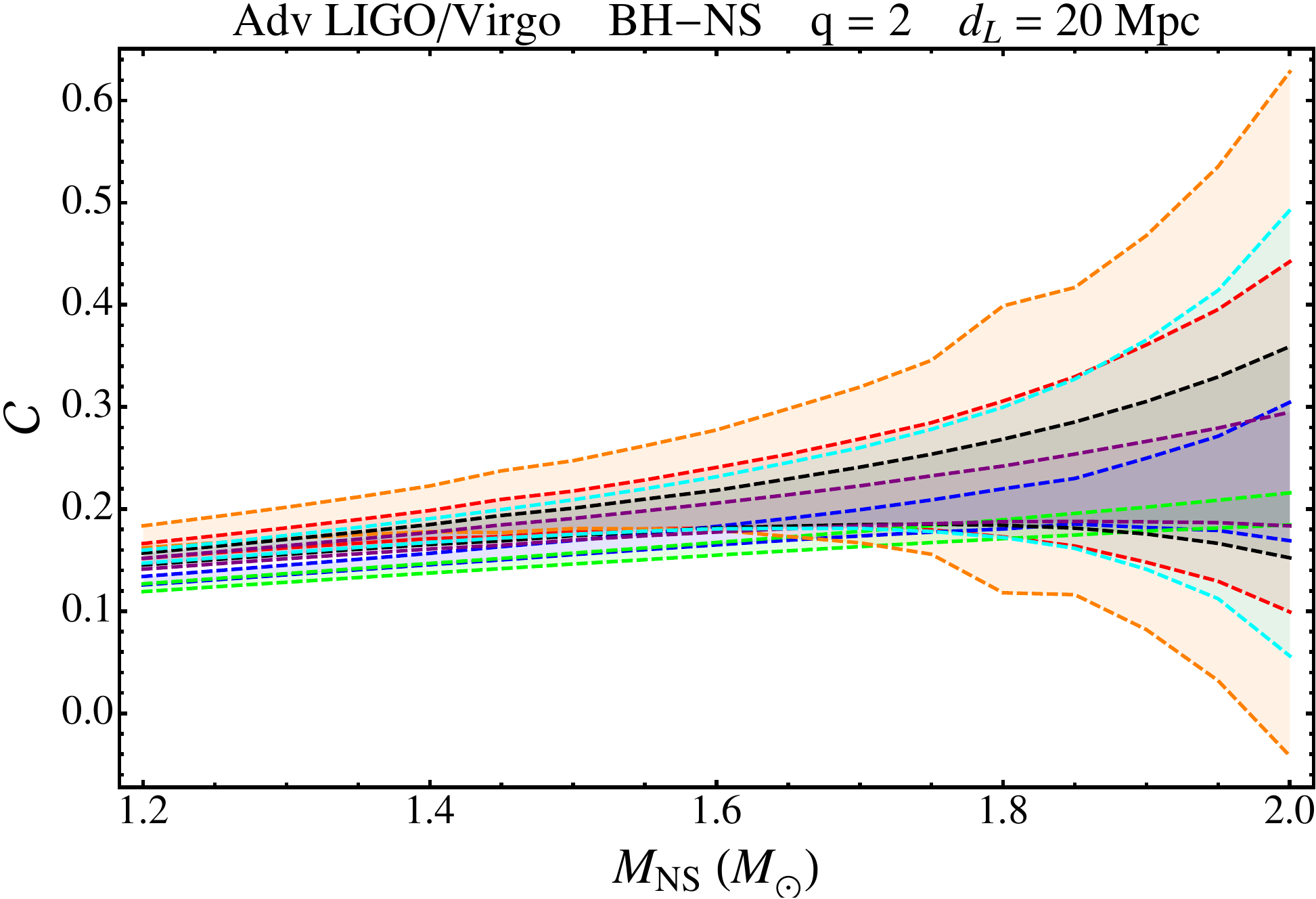}\\ \vspace{0.5cm}
\includegraphics[width=8.1cm]{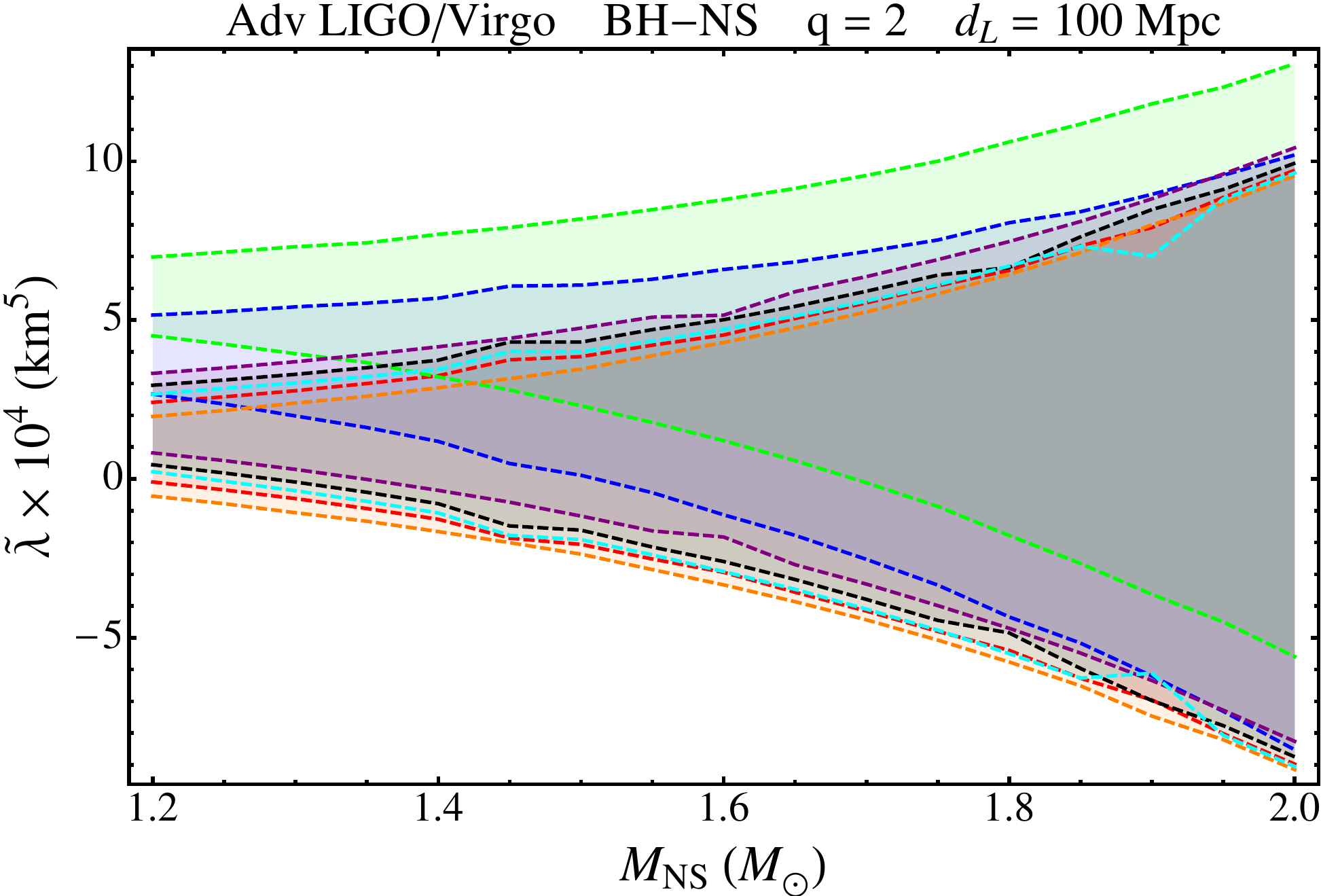}
\includegraphics[width=8.1cm]{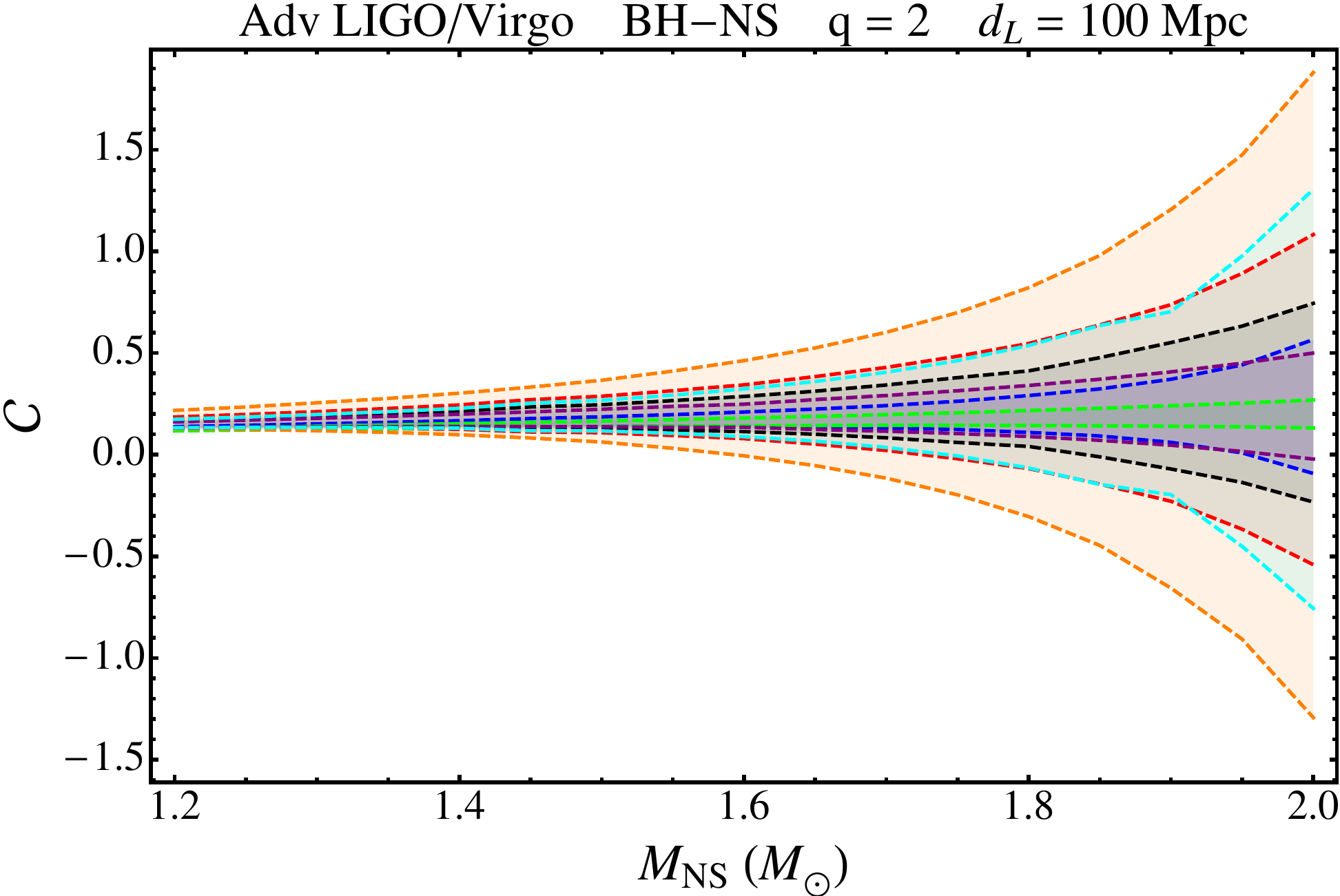} \caption{The quantities $\tilde\lambda\pm
  \sigma_{\tilde\lambda}$ (left column) and ${\cal C}_{\lambda}\pm \sigma_{{\cal
      C}_{\lambda}}$ (right column) are plotted as functions of the NS mass, at
  different luminosity distances $d_{\tn{L}}$, for NS-NS binaries and for BH-NS
  systems with mass ratio $q=2$. The parameter errors $\sigma_{\tilde\lambda}$
  and $\sigma_{{\cal C}_{\lambda}}$ are evaluated for AdvLIGO/Virgo.  Different
  bands correspond to different NS EoS.}
\label{errorsAL}
\end{figure*}

\begin{figure*}[htp]
\centering
\includegraphics[width=8.cm]{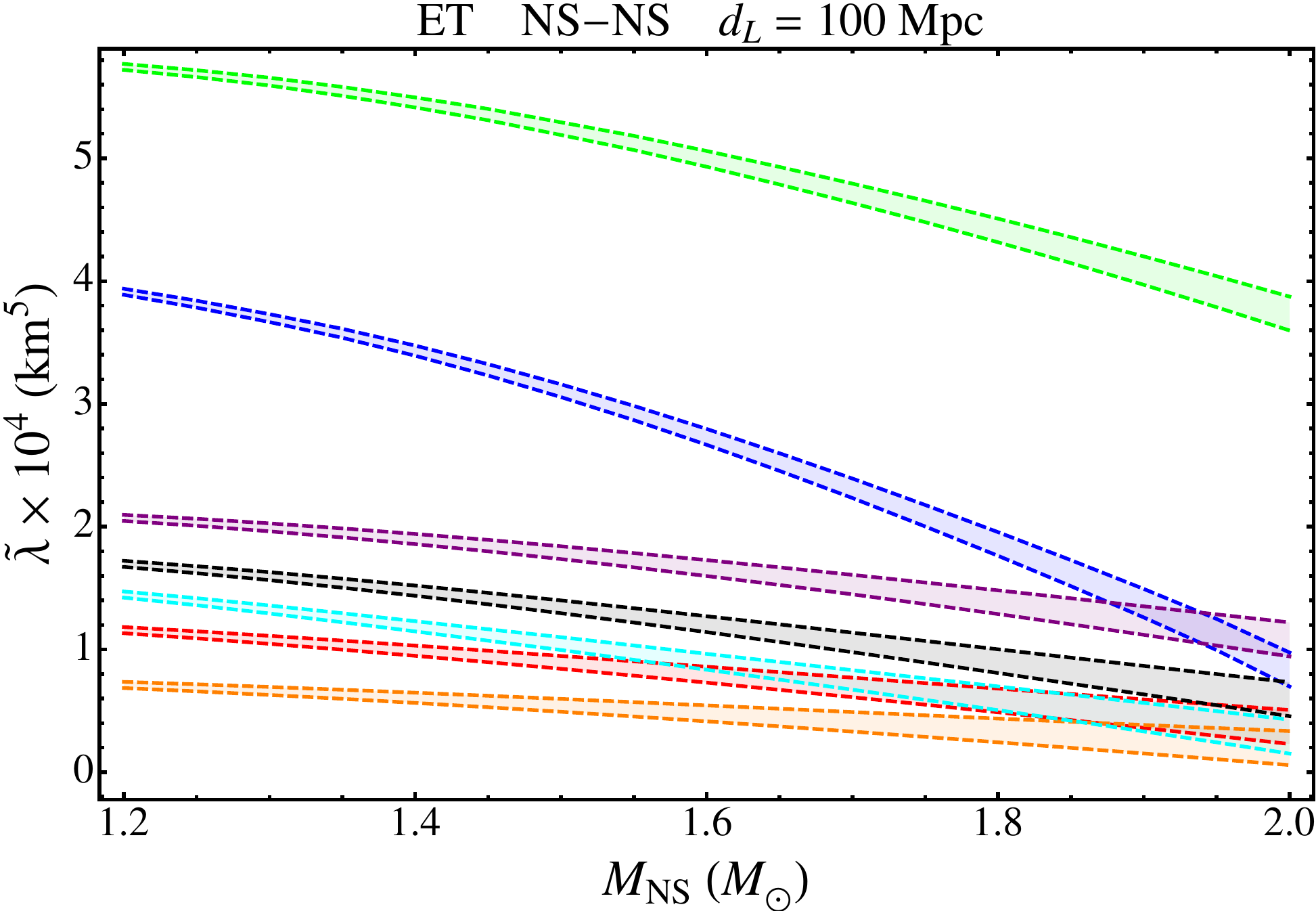}
\includegraphics[width=8.1cm]{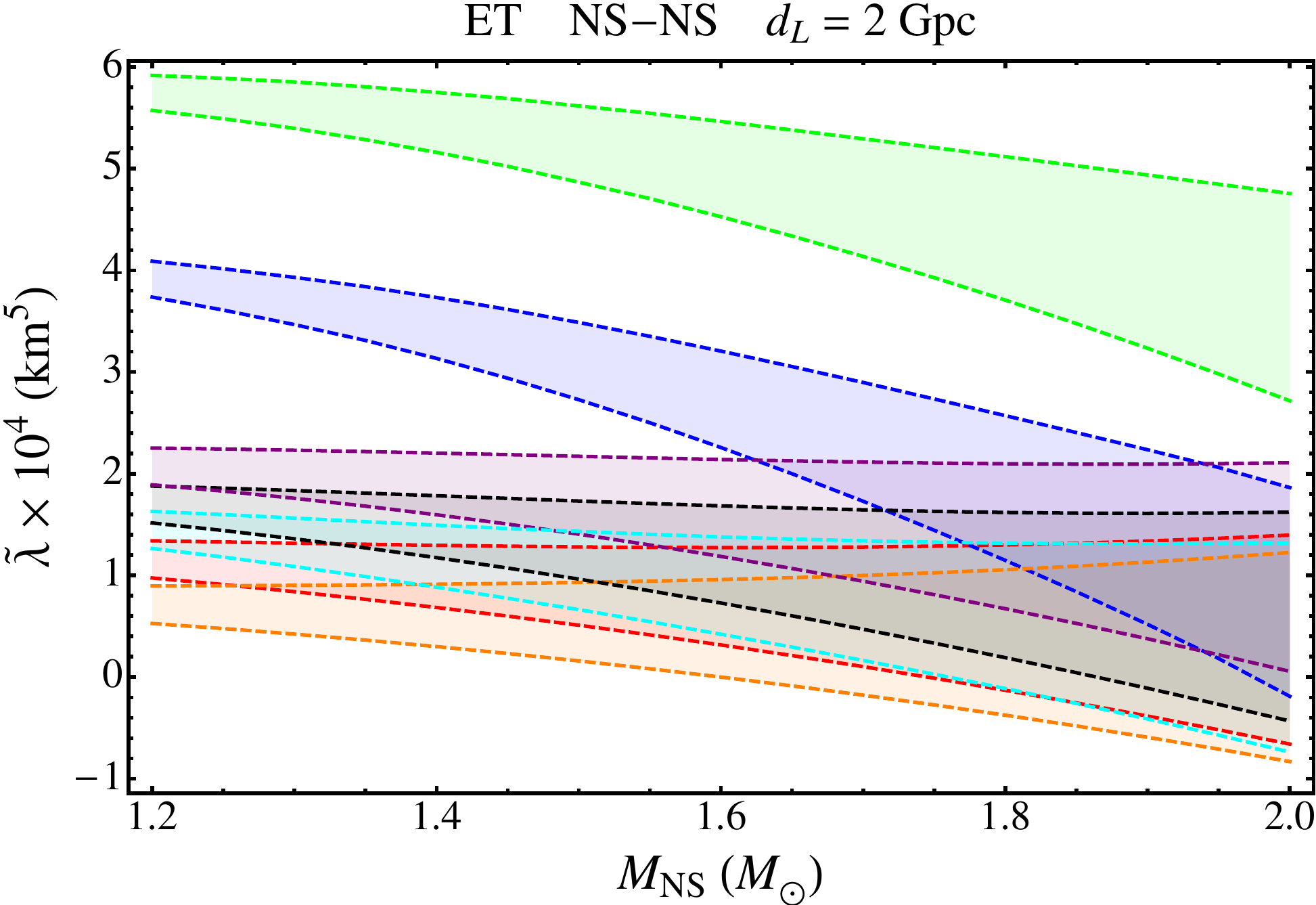}\\
\vspace{0.3cm}
\includegraphics[width=8cm]{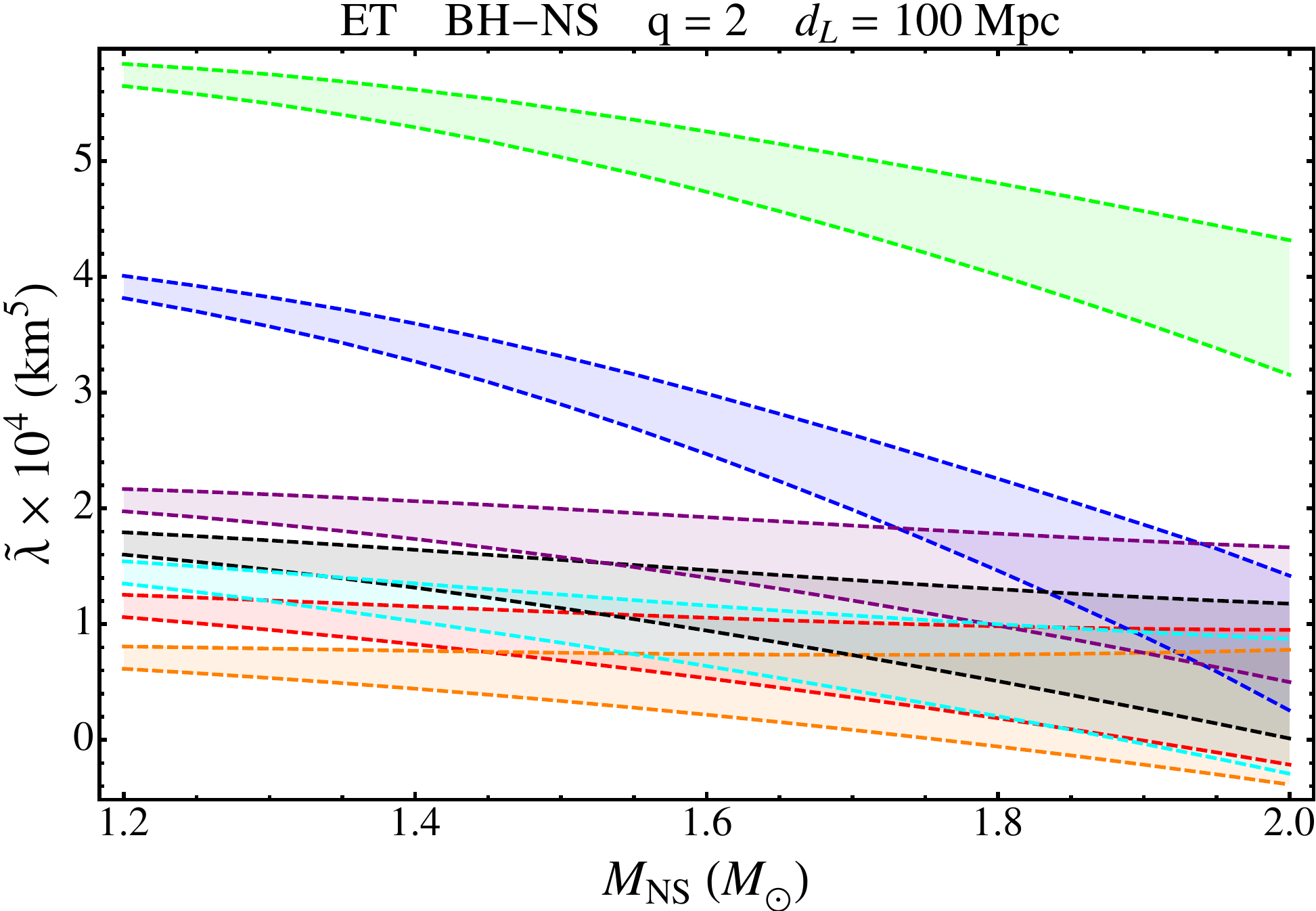}
\includegraphics[width=8.1cm]{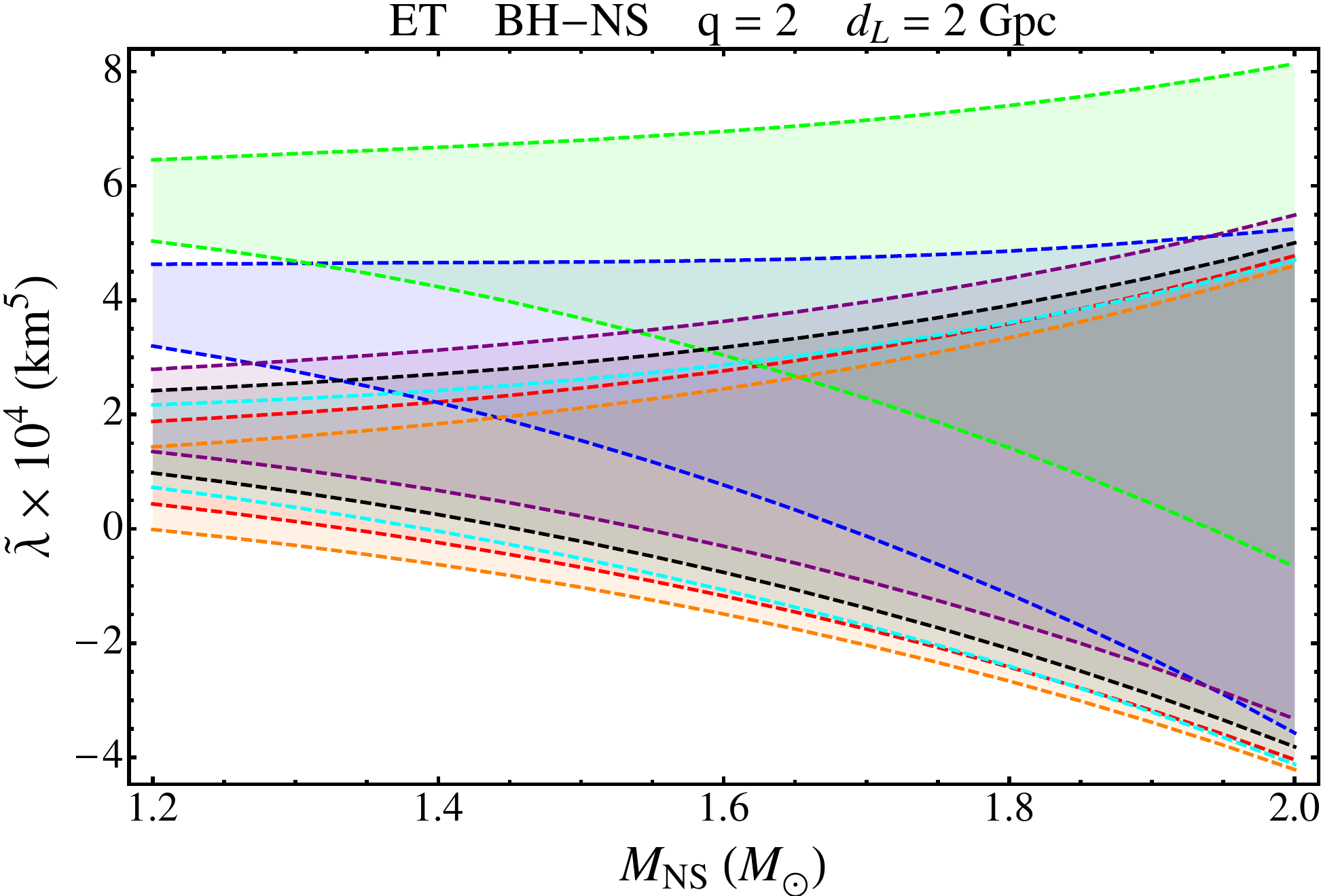}\\
\vspace{0.3cm}
\includegraphics[width=8.cm]{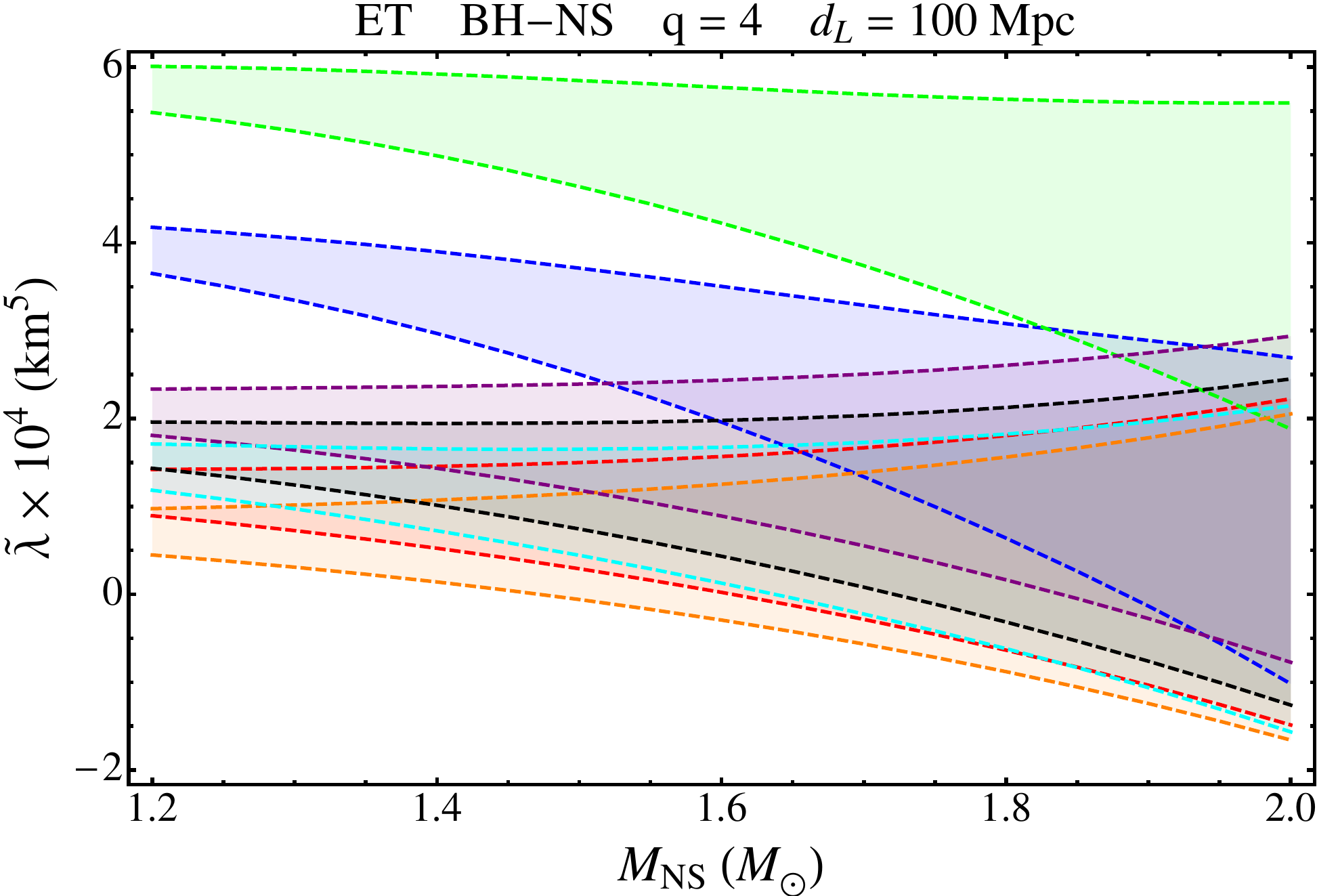}
\includegraphics[width=8.1cm]{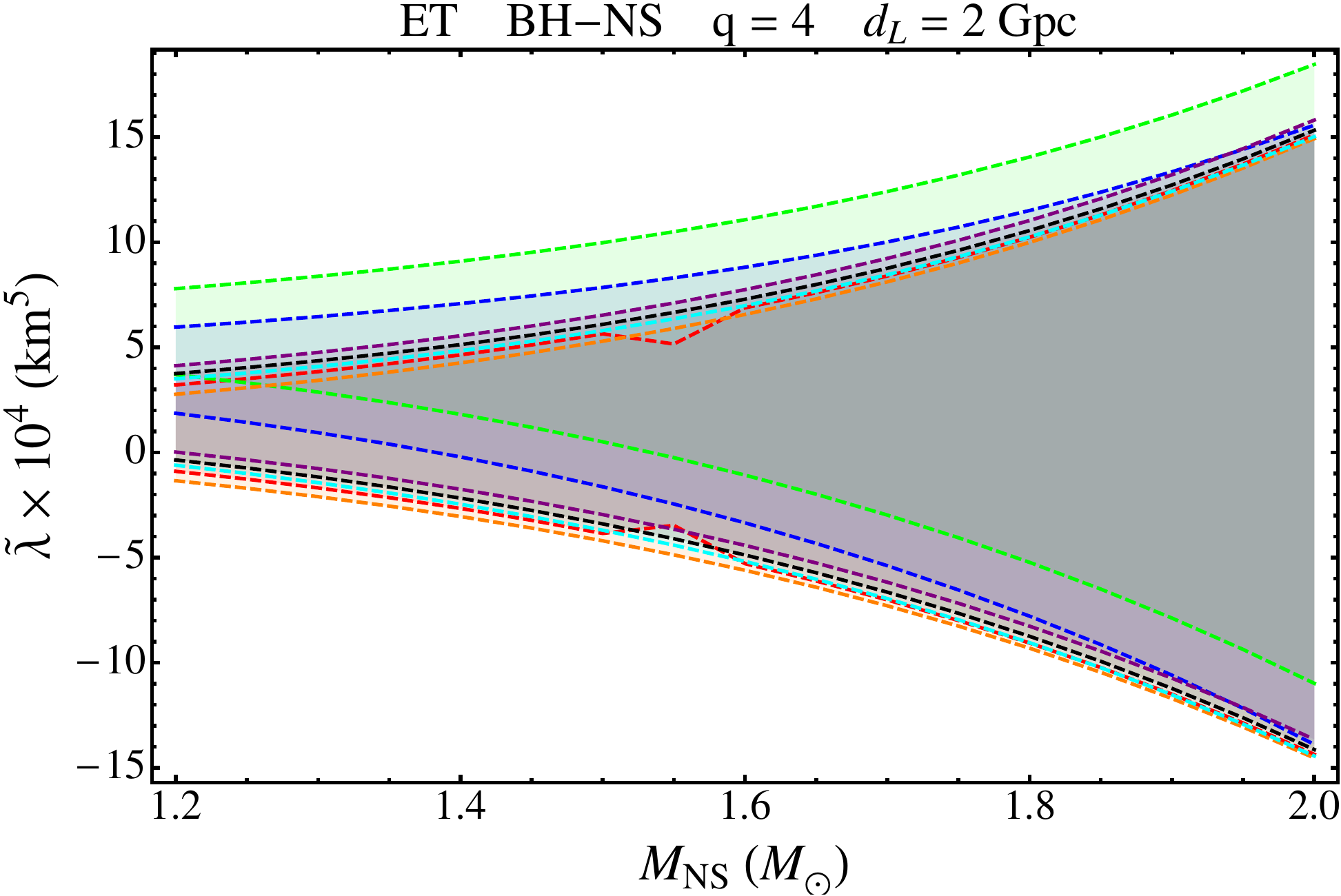}
\caption{ The quantity $\tilde\lambda \pm \sigma_{\tilde\lambda}$ is plotted as
  a function of the NS mass, for sources at luminosity distance $d_{\tn{L}}=100$
  Mpc (left panels) and $d_{\tn{L}}=100$ Gpc (right panels), for NS-NS systems
  and for BH-NS systems with $q=2$ and $q=4$, and for the same EoS considered in
  Fig.~\ref{errorsAL}. The errors are evaluated for the Einstein Telescope.}
\label{errorsET}
\end{figure*}

We consider NS-NS binaries with mass ratio $q=1$, and BH-NS binaries with mass
ratio $q=2,4$; in the first case both stars are modeled with the same equation
of state. The error on the tidal deformability is directly computed by means of
a $5\times 5$ Fisher Matrix for the set of parameters
$\boldsymbol\theta=(t_{c},\phi_{c},\ln{\cal M},\ln\nu,\tilde\lambda)$. The
uncertainty on ${\cal C}$ is obtained in terms of the uncertainty on
$\tilde\lambda$, using the fit (\ref{ILQ}) (see Eq.~(\ref{properr})).  For each
detector we consider prototype binaries at fixed
distances: \begin{itemize} \item For the Advanced detectors, we analyze systems
  at luminosity distance $d_{\tn{L}}=(20,100)$Mpc.  In this case the $S_{h}(f)$
  is taken to be the \texttt{ZERO\_DET\_high\_P} anticipated sensitivity curve
  \cite{zerodet}.  \item For ET we consider binaries at distance up to
  $d_{\tn{L}}=1$ Gpc, and use the noise spectral density given by the analytic
  fit (\ref{ETLfit}).  \end{itemize} We employ the standard PN
template \begin{equation} h_{\textnormal{PN}}={\cal A}_\textnormal{3PN}(f)\
  e^{i(\psi_\textnormal{PP}+\psi_\textnormal{T})}\,.  \end{equation} All
quantities are suitably redshifted.  The integrations are performed in the
frequency range $[f_{\tn{min}},f_{\tn{ISCO}}]$, where $f_{\tn{min}}$ is set to
20 Hz for AdvLIGO/Virgo and 10 Hz for ET, and
$f_{\tn{ISCO}}=(6^{3/2}m)^{-1}$. The results are shown in Figs.~\ref{errorsAL}
and \ref{errorsET}, where we plot the intervals
$\tilde\lambda\pm\sigma_{\tilde\lambda}$ (left panels of Fig.~\ref{errorsAL},
and Fig.~\ref{errorsET}) and $\cal C\pm\sigma_{\cal C}$ (right panels of
Fig.~\ref{errorsAL}) as functions of the NS mass for AdvLIGO/Virgo and ET,
respectively. In the headline of each panel we indicate the detector, the mass
ratio and the source distance.

Let us start discussing the results for AdvLIGO/Virgo shown in
Fig.~\ref{errorsAL}.  In the upper, left panel, we consider NS-NS binaries as
close as $20$ Mpc; we see that, for the softer EoS (WFF1, APR4, SLy4, ENG, MPA1)
the tidal deformability $\tilde\lambda$ is weakly dependent on the NS mass; for
stiffer EoS (H4, MS1) $\tilde\lambda$ exhibits a more pronounced dependence on
$M_{\tn{NS}}$; in addition, for each EoS $\tilde\lambda$ varies in ranges
(1\,$\sigma$ intervals) which are more separated for stiffer EoS, and for masses
smaller that $\sim 1.8~M_\odot$.  Conversely, the compactness ${\cal{C}}$ shown
in the right, upper panel does not seem to be a good indicator of the EoS.

If the sources are farther away, say at $100$ Mpc as shown in the middle panels,
chances to discriminate among the EoS decrease, because the 1\,$\sigma$
intervals become larger, and only when the EoS is stiff and the mass is lower
than $\sim 1.5~M_\odot$, the quantity $\tilde\lambda$ can be used as EoS
indicator.  In the four lower panels of Fig.~\ref{errorsAL} we consider BH-NS
binaries with $q=2$ and $d_{\tn{L}}=20$ and $d_{\tn{L}}=100$ Mpc, respectively.
We see that advanced detectors will be able to extract some information on the
EoS from $\tilde\lambda$ only if the source is very close and the NS mass is
lower than $\sim 1.5~M_\odot$, but it should be reminded that the rate of BH-NS
coalescence is much smaller than that of NS-NS, therefore to obtain a reasonable
detection rate one should have access to a much larger volume space, as that
allowed by the third generation detectors like ET. When $q>2$, the chances to
discriminate among the EoS are even smaller.  In any event, comparing the right
and the left panels of Fig.~\ref{errorsAL} we can conclude that $\tilde\lambda$
is the parameter to use if we want to discriminate among different EoS, although
its effectiveness decreases with the source distance.  Since this is true also
for third generation detectors, in the following we shall consider only the
plots for $\tilde\lambda$.  They are shown in Fig.~\ref{errorsET} for the
detector ET.  As noted before, the possibility to discriminate between different
EoS decreases as the distance increases; however, if the source is a NS-NS
system (upper panels) it remains acceptable even when $d_{\tn{L}}$ is as large
as the ET horizon distance, which is estimated to be about $2$ Gpc (right
panel). If the source is close (left panel $d_{\tn{L}}=100$ Mpc), the error on
$\tilde\lambda$ is very small and, unless the NS masses are close to the maximum
observed mass there are good chances to identify at least the class to which the
EoS belongs.  The remaining four panels of Fig.~\ref{errorsET} refer to BH-NS
systems at a distance of $100$ Mpc (left) and of $2$ Gpc (right).  The middle
panels refer to systems with mass ratio $q=2$, the lower panels to more
realistic systems with $q=4$.  We see that there is no way to give reliable
information on the EoS if this kind of source is at cosmological distance,
regardless of the value of $q$. For closer systems, and if the mass is smaller
than a certain value, we could infer if the EoS is soft or stiff; for instance
this could be possible for sources at $d_{\tn{L}}=100$ Mpc if $M_{\tn{NS}}$ is
smaller than $\sim 1.7~M_\odot$, for systems with $q=2$, and if $M_{\tn{NS}}$ is
smaller than $\sim 1.5~M_\odot$, for systems with $q=4$.

\section{Concluding remarks}\label{sec:concl}

In this work we have discussed how to extract information on the EoS of matter
in a neutron star interior, using the gravitational wave signal emitted in NS-NS
or BH-NS coalescence.  In a few years the second generation of interferometric
detectors AdvLIGO/Virgo should detect GW signals emitted by these sources, and
chances of detection will significantly increase with the third generation
detectors.  The information on the NS EoS is encoded in the signal emitted
during the latest phases of inspiralling before merging by means of two
quantities: the NS tidal deformability and, when tidal disruption occurs before
merger, the cut-off frequency.  Assuming that a signal emitted by one or more of
such systems is detected, we have evaluated the accuracy with which different
parameters can be determined, using different data analysis strategies based on
the Fisher matrix approach and on analytical fits of the relevant quantities
obtained by numerical simulations of the coalescence process.

Our main results can be summarized as follows. 

\begin{itemize}
\item{Using $f_{\tn{cut}}$ to gain information on the NS EoS.}

  We find that, using $f_{\tn{cut}}$ as free parameter in the Fisher matrix
  approach and the first proposed strategy, the NS compactness ${\cal C}$ can be
  estimated with a relative error of the order of $3\%-4\%$, and that its
  estimate has a weak dependence on $d_{\tn{L}}$, being the error dominated by
  the error on the fit ${\cal C}(\lambda)$ (see Eq.~(\ref{properr})).  The
  situation does not change if we use the second strategy, i.e.  if we do not
  include $f_{\tn{cut}}$ as unknown parameter in the Fisher matrix, and express
  the waveform template only in terms of the tidal deformation, using the fits
  $f_{\tn{cut}}({\cal{C}})$ and ${\cal{C}}(\lambda)$.  Thus, if our goal is to
  estimate ${\cal C}$, the use of $f_{\tn{cut}}$ in the analysis is
  ineffective. However, if the goal is to estimate the tidal deformability
  $\tilde\lambda$, it is better to use the second strategy because the error
  reduces up to $30\%$ for more distant sources.  To make this analysis we have
  used the same models used in \cite{KST10} to obtain the fit
  $f_{\tn{cut}}({\cal{C}})$, i.e.  non spinning BH-NS binaries with mass ratio
  $q=2$, with the neutron star modeled using a set of piecewise polytropes.

  This study could, and should, be extended to more general BH-NS binaries, with
  different values of the mass ratio and different EoS, and including the BH
  spin.  The extension to spinning BHs is important because when the mass ratio
  is larger, or the EoS is less deformable, NS tidal disruption can occur only
  if the BH is rapidly rotating. Such extension will be possible only when fully
  relativistic, numerical simulations will extend the domain of validity of the
  fit $f_{\tn{cut}}({\cal C})$.  In addition, it should be stressed that the
  presence of a frequency cutoff in the GW signal indicates that the NS has been
  tidally disrupted before merging with the companion; if the torus of dense
  matter which subsequently forms has sufficiently high mass, a Short Gamma-Ray
  Burst could be powered.  Therefore, estimating $f_{\tn{cut}}$ as accurately as
  possible is also important.

\item{Comparing $\tilde\lambda$ and ${\cal{C}}$ as EoS indicators.}
 
  We have discussed and compared the effectiveness of $\tilde\lambda$ and
  ${\cal{C}}$ in discriminating between different EoS.  To this aim, we have
  considered NS-NS and NS-BH binaries, modeled using a large set of EoS, to be
  detected by $2^{nd}$ and $3^{rd}$ generation interferometers. We find that
  $\tilde\lambda$ is much better than ${\cal{C}}$ to constrain the NS equation
  of state.

  For sources at distance $d_{\tn{L}}\lesssim100$ Mpc, a signal emitted by NS-NS
  binaries detected by Advanced LIGO/Virgo would allow to discriminate between
  different EoS, if $M_{\tn{NS}}\lesssim1.5~M_\odot$. If the signal is detected
  by ET, different EoS will be discerned even for larger distances and larger NS
  masses. Conversely, it is very unlikely that BH-NS binaries will allow us to
  discriminate between different EoS: this could happen only in the unlikely
  case of a signal detected by ET from a system with
  $M_{\tn{NS}}\lesssim1.3~M_\odot$ and $q\lesssim2$.
\end{itemize}

Other improvement of this strategy that should be pursued are:
\begin{itemize}
\item Use of more realistic gravitational wave templates, aimed at reproducing
  in detail the full merger of BH-NS binaries \cite{PBKS13}.
\item Correlation of other binary parameters related to the neutron star
 structure.  For instance, it has recently been shown that the peak frequency
 of the post merger GW signal from NS-NS mergers is strongly related with the
 NS radius \cite{BJHS12}.
\item Use of Bayesian inference for GW parameter estimation based on Markov
 Chain Monte Carlo algorithms \cite{BJ12,BSD96}.
\end{itemize}

\appendix

\section{Parameter estimation: basic Theory}\label{app:Fisher}
In this section we briefly recall the key points of parameter estimation
theory. Given a signal of the form $h(t,\boldsymbol\theta)$, we want to extract
the physical parameters $\boldsymbol\theta$ and estimate the errors
$\Delta\boldsymbol\theta=\boldsymbol\theta -\hat{\boldsymbol\theta}$, were we
assume $\boldsymbol{\hat\theta}$ to be the true values of the parameters. To
this aim we have to compute $p(\boldsymbol\theta\vert s)$, namely the
probability of measuring the value $\boldsymbol\theta$, given the detector
output $s(t)$
\begin{equation}
s(t)=h(t,\boldsymbol\theta)+n(t)\ ,
\end{equation}
where $n(t)$ represents the detector noise, which is assumed to be stationary. It has 
been shown that \cite{CF94}
\begin{equation}\label{prob}
p(\boldsymbol\theta\vert s)\propto p^{(0)}(\boldsymbol\theta)e^{-\frac{1}{2}(h(\boldsymbol\theta)
-s\vert h(\boldsymbol\theta)-s))}\ ,
\end{equation}
where $p^{(0)}(\boldsymbol\theta)$ is the prior probability on the physical set
of parameters and the constant of normalization is independent of
$\boldsymbol\theta$. We define the inner product $(\cdot \vert \cdot)$ as
\begin{equation}\label{inner}
(g\vert h)=2\int_{-\infty}^{\infty}\frac{\tilde h(f)\tilde g^{\star}(f)+\tilde h^{\star}(f)\tilde g(f)}{S_{h}(f)}df\ ,
\end{equation} 
where $\tilde h(f)$ is the Fourier transform of $h(t)$ and $^{\star}$ denotes
complex conjugation. The integration range can be different from
$[-\infty,+\infty]$, if specified. $S_{n}(f)$ is the noise spectral density
(PSD) of the detector considered.  For a given measurement the values of the
source parameters can be estimated as those which maximize the probability
distribution (\ref{prob}): this is the so called {\it maximum-likelihood
  estimator}. Moreover, we define the signal-to-noise ratio $\rho$, such that
\begin{equation}
\rho^{2}=\left(h\vert h\right)=4\int_{0}^{\infty}\frac{\vert \tilde h(f)\vert^{2}}{S_{h}(f)}df\ ,
\end{equation}
evaluated at $\boldsymbol\theta =\hat{\boldsymbol\theta}$. In the following we will consider 
the limit of large signal to noise ratio (SNR), in which $p(\boldsymbol\theta\vert h)$ is sharply peaked around the true values of 
the source parameters. It can be easily proved that in this limit 
\begin{equation}\label{prob1}
p(\boldsymbol\theta\vert s)\propto p^{(0)}(\boldsymbol\theta)e^{-\frac{1}{2}\Gamma_{ab}\Delta\theta^{a}\Delta\theta^{b}}\ ,
\end{equation}
where 
\begin{equation}\label{FisherM}
\Gamma_{ab}=\left(h_{,a}\vert h_{,b}\right)
\end{equation}
evaluated at $\boldsymbol\theta=\hat{\boldsymbol\theta}$, is the {\it Fisher
  information matrix} \cite{PW95}. This allows to define the variance-covariance
matrix $\Sigma^{ab}$ as\footnote{Angular brackets denote average on the
  probability distribution defined by Eq.~(\ref{prob1}).}
\begin{equation}
\Sigma^{ab}=\langle \Delta\theta^{a}\Delta\theta^{b}\rangle=\left(\boldsymbol\Gamma^{-1}\right)^{ab}\ ,
\end{equation}
where $\boldsymbol\Gamma^{-1}$ is the inverse of Fisher matrix. In this way we
can define the error associated to the parameter $\theta^{a}$ as
\begin{equation}\label{error}
\sigma_{a}=\langle (\Delta\theta^{a})^{2}\rangle^{1/2}=\sqrt{\Sigma^{aa}}\ ,
\end{equation}
and the correlation coefficient between $\theta^{a}$ and $\theta^{b}$ as
\begin{equation}\label{corr}
c_{ab}=\frac{\langle \Delta\theta^{a}\Delta\theta^{b}\rangle}{\Sigma^{aa}\Sigma^{bb}}=
\frac{\Sigma^{ab}}{\sqrt{\Sigma^{aa}\Sigma^{bb}}}\ ,
\end{equation}
with $c_{ab}\in[-1,1]$.  Eq.~(\ref{prob1}) holds whether
$p^{(0)}(\boldsymbol\theta)$ is uniform around $\hat{\boldsymbol\theta}$ or
not. In the first case, the probability distribution for the parameters
takes a Gaussian form. Otherwise $p(\boldsymbol\theta\vert s)$ doesn't represent
the maximum-likelihood estimate, and it may not be Gaussian.  In the special
case when $p^{(0)}$ is Gaussian, namely
\begin{equation}
p^{(0)}(\boldsymbol\theta)\propto e^{-\frac{1}{2}\Gamma_{ab}^{(0)}(\theta^{a}-\bar\theta^{b})(\theta^{b}-\bar\theta^{b})}\ ,
\end{equation}
the probability distribution $p(\boldsymbol\theta\vert s)$ is Gaussian with
covariance matrix given by
\begin{equation}
\boldsymbol\Sigma=(\boldsymbol\Gamma+\boldsymbol\Gamma^{(0)})^{-1} \ .
\end{equation}
We note that in general $p(\boldsymbol\theta\vert s)$ is peaked around
$\langle\boldsymbol\theta\rangle$, which is, in general, different from
$\bar{\boldsymbol\theta}$ and $\hat{\boldsymbol\theta}$.

\acknowledgments
We would like to thank M. Fortin and L. Pagano for useful discussions.
A.M. is supported by a ``Virgo EGO Scientific Forum'' (VESF) grant.

\bibliographystyle{h-physrev4}

\begin{thebibliography}{99}
\bibitem{LIGOVirgo} \url{http://www.ligo.caltech.edu}, \url{http://www.ego-gw.it}
\bibitem{ET} \url{http://www.et-gw.eu/}.
\bibitem{H08} T.~Hinderer, Astrophys.\ J.\ {\bf 677}, 1216 (2008); {\it ibid.}, {\bf 697}, 964 (2009).
\bibitem{FH08} E.E.~Flanagan, T.~Hinderer, Phys. Rev. {\bf D77}, 021502 (2008).
\bibitem{Val200} M.~Vallisneri, Phys. Rev. Lett. {\bf 84}, 3519 (2000).
\bibitem{FGP10} V.~Ferrari, L.~Gualtieri and F.~Pannarale, Phys. Rev. D {\bf 81}, 064026 (2010).
\bibitem{HLLR10} T.~Hinderer, B.~D.~Lackey, R.~N.~Lang and J.~S.~Read, Phys.\ Rev.\ D {\bf 81}, 123016 (2010).
\bibitem{KST10} K.~Kyutoku, M.~Shibata, K.~Taniguchi, Phys. Rev. D. {\bf 82}, 044049 (2010).
\bibitem{DFKOT10}  M.~D.~Duez, F.~Foucart, L.~E.~Kidder, C.~D.~Ott and S.~A.~Teukolsky, 
Class.\ Quant.\ Grav.\  {\bf 27}, 114106 (2010).
\bibitem{PTR11} F.~Pannarale, L.~Rezzolla, F.~Ohme and J.~S.~ Read, Phys. Rev. D. {\bf 84}, 
104017 (2011).
\bibitem{LKSBF11} B.~D.~Lackey, K.~Kyutoku, M.~Shibata, P.R.~Brady, J.L.~Friedman, Phys. Rev. 
{\bf D} 85 044061 (2012).
\bibitem{DNV12} T.~Damour, A.~Nagar, L.~Villain, Phys. Rev. {\bf D} 85, 123007 (2012).
\bibitem{Ral13} J.~S.~Read, L.~Baiotti, J.~D.~E.~Creighton, J.~L.~Friedman, B.~Giacomazzo, 
K.~Kyutoku, C.~Markakis and L.~Rezzolla {\it et al.}, Phys.\ Rev.\ D {\bf 88}, 044042 (2013).
\bibitem{MCFGP13}  A.~Maselli, V.~Cardoso, V.~Ferrari, L.~Gualtieri and P.~Pani, Phys.\ Rev.\ D {\bf 88}, 023007 (2013).
\bibitem{NPP92} R.~Narayan, B.~Paczynski and T.~Piran, Astrophys.\ J.\  {\bf 395}, L83 (1992).
\bibitem{BBM13} I.~Bartos, P.~Brady and S.~Marka, Class.\ Quant.\ Grav.\  {\bf 30}, 123001 (2013).
\bibitem{LP07} J.~M.~Lattimer and M.~Prakash, Phys.\ Rept.\  {\bf 442}, 109 (2007).
\bibitem{YY13}  K.~Yagi and N.~Yunes, Science {\bf 341}, 365 (2013); Phys.\ Rev.\ D {\bf 88}, 023009 (2013).
\bibitem{DN09} T.~Damour and A.~Nagar, Phys.\ Rev.\ D {\bf 80}, 084035 (2009).
\bibitem{BP09} T.~Binnington and E.~Poisson, Phys.\ Rev.\ D {\bf 80}, 084018 (2009).
\bibitem{VFH11} J.~Vines, E.E.~Flanagan, T.~Hinderer, Phys. Rev. {\bf D} 83, 084051 (2011).
\bibitem{BDF12} D.~Bini, T.~Damour, G.~Faye, Phys. Rev. D {\bf 85} 124034, (2012).
\bibitem{CF94} C.~Cutler and E.~E.~Flanagan, Phys.\ Rev.\ D {\bf 49}, 2658 (1994).
\bibitem{BSD96}  R.~Balasubramanian, B.~S.~Sathyaprakash and S.~V.~Dhurandhar, Phys.\ Rev.\ D {\bf 53}, 3033 (1996).
\bibitem{DABV13} W.~Del Pozzo, T.~G.~F.~Li, M.~Agathos, C.~V.~D.~Broeck and S.~Vitale,  
Phys.\ Rev.\ Lett. {\bf 111}, 071101 (2013).
\bibitem{SKYT09} M.~Shibata, K.~Kyutoku, T.~Yamamoto and K.~Taniguchi, Phys. Rev. D {\bf 79}, 044030 (2009); 
{\bf 85}, 127502(E) (2012).
\bibitem{KOST11} K.~Kyutoku, H.~Okawa, M.~Shibata and K.~Taniguchi, Phys. Rev. D {\bf 84}, 064018 (2011).
\bibitem{N05}  R.~Narayan, New J.\ Phys.\  {\bf 7}, 199 (2005).
\bibitem{BDGNR}  L.~Baiotti, T.~Damour, B.~Giacomazzo, A.~Nagar and L.~Rezzolla, Phys.\ Rev.\ Lett.\  {\bf 105}, 261101 (2010);
Phys.\ Rev.\ D {\bf 84}, 024017 (2011).
\bibitem{BTB12}  S.~Bernuzzi, M.~Thierfelder and B.~Bruegmann, Phys.\ Rev.\ D {\bf 85}, 104030 (2012).
\bibitem{BNTB12} S.~Bernuzzi, A.~Nagar, M.~Thierfelder and B.~Bruegmann, Phys.\ Rev.\ D {\bf 86}, 044030 (2012).
\bibitem{HKS13}  K.~Hotokezaka, K.~Kyutoku and M.~Shibata, Phys.\ Rev.\ D {\bf 87}, 044001 (2013).
\bibitem{RLG13}  D.~Radice, L.~Rezzolla and F.~Galeazzi, Mon.\ Not.\ Roy.\ Astron.\ Soc. {\it published online},
 doi:10.1093/mnrasl/slt137 (2013).
\bibitem{DIS00} T.~Damour, B.~R.~Iyer and B.S.~Sathyaprakash, Phys. Rv. D {\bf 62}, 084036 (2000).
\bibitem{DNT11} T.~Damour, A~.Nagar, M.~Trias, Phys. Rev. D {\bf 83}, 024006 (2011).
\bibitem{B06} L.~Blanchet, living Rev. Relativity {\bf 9}, 4 (2006).
\bibitem{MR11} C.~Messenger and J.~Read, Phys. Rev. Lett. 108, 091101 (2012).
\bibitem{Planck} Planck Collaboration, arXiv:1303:5062 (2013).
\bibitem{PW95} E.~Poisson and C.~M.~Will, Phys. Rev. D {\bf 52}, 2 (1995).
\bibitem{SC09} B.S.~Sathyaprakash and B.F.~Schultz, Living Reviews in Relativity {\bf 12}, 2 (2009).
\bibitem{RMSUCF09} J.~Read et al., Phys. Rec. D {\bf 79}, 124033 (2009).
\bibitem{VF10} J.~Vines, E.~Flanagan, gr-qc: 1009.4919 (2010).
\bibitem{LOB08} L.~Lindblom, B.J.~Owen and D.A.~Brown, Phys. Rev. D {\bf 78}, 124020 (2008).
\bibitem{APR4} A.~Akmal, V.R.~Pandharipande, and D.G.~Ravenhall, Phys.
Rev. C {\bf 58}, 1804 (1998).
\bibitem{WFF1} R.B.~Wiringa, V.~Fiks, and A.~Fabrocini, Phys. Rev. C
{\bf 38}, 1010 (1988).
\bibitem{MPA1} H.~Muther, M.~Prakash, and T.L.~Ainsworth, Phys. Lett. B ({\bf 199}), 
469 (1987).
\bibitem{ENG} L.~Engvik {\it et al.}, Astrophys. J. {\bf 469}, 794 (1996). 
\bibitem{SLy4} F.~Douchin and P.~Haensel, Astron. Astrophys. {\bf 380}, 151 (2001).
\bibitem{MS1} H.~Muller and B.D.~Serot, Nucl. Phys. {\bf A606}, 508 (1996).
\bibitem{H4} B.D.~Lackey, M.~Nayyar, and B.J.~Owen, Phys. Rev. D {\bf 73}, 024021 (2006).
\bibitem{zerodet} D.~Shoemaker, \texttt{https://dcc.ligo.org/cgi-bin/ \\ DocDB/ShowDocument?docid=2974}.	
\bibitem{PBKS13} F.~Pannarale,E.~Berti, K.~Kyutoku and M.~Shibata,  Phys. Rev. D {\bf 88}, 084011 (2013).
\bibitem{BJHS12} A.~Bauswein, H.T.~Janka, K.Hebeler and A.~Schwenk, Phys. Rev. D {\bf 86}, 063001 (2012). 
\bibitem{BJ12}  A.~Bauswein and H.~-T.~.Janka, Phys.\ Rev.\ Lett.\  {\bf 108}, 011101 (2012).
\end{thebibliography}

\end{document}